\documentclass[aip, reprint, twocolumns, amsmath,amsfonts,amssymb]{revtex4-2}

\usepackage{xcolor}
\usepackage{verbatim}
\usepackage{cancel}
\usepackage{algorithm}
\usepackage[colorlinks=true,
            linkcolor=blue,
            citecolor=blue,
            urlcolor=blue,
            filecolor=blue]{hyperref}

\usepackage{multirow}

\usepackage{graphicx}
\usepackage{amsmath,amsfonts,amssymb}
\usepackage{color}
\usepackage{bm}
\usepackage[mathscr]{eucal}
\usepackage{picture}
\usepackage{float}
\usepackage{subcaption}

\usepackage{graphicx}
\usepackage{amsmath,amsfonts,amssymb}
\usepackage{color}
\usepackage{bm}
\usepackage{threeparttable}
\usepackage[mathscr]{eucal}
\usepackage{picture}

\definecolor{grey}{rgb}{0.7,0.7,0.7}
\newcommand{\black}{\color{black}{}}

\usepackage{fancyvrb}
\usepackage{ulem}

\usepackage[T1]{fontenc}

\definecolor{brown}{RGB}{111,16,50}

\definecolor{purple}{rgb}{0.8,0.0,0.8}
\definecolor{pink}{rgb}{1.,0.5,0.5}

\newcommand{\ctext}[1]{\raise0.2ex\hbox{\textcircled{\scriptsize{#1}}}}

\begin{document}
\title{
Classically Driven Hybrid Quantum Algorithms with Sequential Givens Rotations for Reduced Measurement Cost}
\author{Benjamin Mokhtar}
\affiliation{Graduate School of Engineering and Science, Shibaura Institute of Technology, 3-7-5 Toyosu, Koto-ku, Tokyo 135-8548 Japan}
\author{Noboru Inoue}
\affiliation{Graduate School of System Informatics, Kobe University, 1-1 Rokkodai-cho, Nada-ku, Kobe, Hyogo 657-8501 Japan}

\author{Takashi Tsuchimochi}
\email{tsuchimochi@gmail.com}
\affiliation{College of Engineering, Shibaura Institute of Technology, 3-7-5 Toyosu, Koto-ku, Tokyo 135-8548 Japan}
\affiliation{Institute for Molecular Science, 38 Nishigonaka, Myodaiji, Okazaki 444-8585 Japan}

\begin{abstract}
Quantum algorithms for electronic-structure simulations are actively being developed, yet many hybrid quantum–classical approaches are bottlenecked by the measurement overhead associated with large molecular Hamiltonians. Here we introduce a diagonalization-driven framework that progressively drives the electronic Hamiltonian toward a (block-)diagonal form in the Slater-determinant basis using sequential Givens rotations. In contrast to Schr\"odinger-picture methods that variationally optimize a wave function, our approach adopts a Heisenberg-picture viewpoint: the Hamiltonian is iteratively transformed, and rotation angles are determined classically from low-dimensional effective blocks, reducing the quantum workload to a small, fixed set of matrix-element measurements per iteration. Candidate generators are estimated via approximate Baker–Campbell–Hausdorff updates with truncation and cumulant-based approximations that control Hamiltonian growth, complemented by stochastic selection to avoid stagnation. We further introduce an angle-merging procedure that reduces circuit depth by consolidating repeated small-angle rotations. We benchmark the framework on N$_2$ and strongly correlated hydrogen systems, assessing convergence behavior, residual-structure diagnostics, measurement–accuracy trade-offs, circuit costs, and robustness under finite sampling.
\end{abstract}
\maketitle

\section{Introduction}
Advances in quantum computing have drawn increasing attention in chemistry, particularly for electronic structure simulations\cite{UniversalQuantumSimulator,SimulatedQuantumComputationofMolecularEnergies}. Despite recent progress, quantum computers remain limited in their applicability to such problems. Current Noisy Intermediate-Scale Quantum (NISQ) hardware suffers from limited number of qubits, short  coherence times, and significant gate errors\cite{Motta_emergingQC,Preskill_2018}. These limitations hinder the accuracy and scalability required for chemically relevant simulations.
Initial quantum approaches to electronic structure, such as the Phase Estimation Algorithm (PEA), demonstrated good potential to obtain exact ground state energies by allowing the direct extraction of eigenphases of the Hamiltonian\cite{kitaev1995quantummeasurementsabelianstabilizer,Abrams1999}. 
PEA is widely regarded as a gold-standard approach for ground-state energy estimation in the fault-tolerant quantum computing (FTQC) regime; however, its requirement of deep quantum circuits and long coherence times renders it impractical for current NISQ devices \cite{OMalley2016}.
To overcome these limitations, hybrid quantum-classical algorithms such as the Variational Quantum Eigensolver (VQE) have been developed~\cite{Peruzzo2014}. VQE reduces circuit depth by leveraging a variational approach in which parameterized quantum circuits prepare trial wave functions, while a classical optimizer iteratively minimizes the expected energy. This strategy is particularly suitable for NISQ hardware and has been widely adopted in quantum computational chemistry\cite{tilly_variational_2022,fedorov_vqe_2022}.

Among the various choices of ans\"atze within the VQE framework, the most conventional is the Unitary Coupled Cluster (UCC), particularly in its singles and doubles form (UCCSD)\cite{bartlett_alternative_1989,kutzelnigg_error_1991,szalay_alternative_1995,taube_2006,cooper_benchmark_2010}. While UCCSD often performs well in weakly correlated regimes\cite{romero_strategies_2018,shen_quantum_2017}, its single-reference character can limit its ability to capture strong static correlations\cite{sokolov_quantum_2020, tsuchimochi_2020}. This has motivated adaptive approaches that construct the ans\"atze dynamically by selecting operators most relevant to the system under study. One prominent example is the Adaptive Derivative-Assembled Pseudo-Trotter Variational Quantum Eigensolver (ADAPT-VQE)~\cite{grimsley_adaptive_2019}, which iteratively grows a compact circuit by selecting operators from a predefined pool based on energy gradients.

Despite advances in ans\"atze design and algorithmic development, a central bottleneck in hybrid quantum-classical algorithms remains the measurement cost associated with estimating expectation values of molecular Hamiltonians and related observables. In the second-quantized representation,  the electronic Hamiltonian contains many fermionic excitation terms, which, after mapping to qubit operators (e.g., Jordan–Wigner\cite{article:JordanWigner1928} or Bravyi–Kitaev\cite{bravyi_fermionic_2002}), become extensive sums of Pauli strings. This proliferation directly increases runtime and amplifies shot noise under finite sampling, posing a major obstacle to scalable quantum simulation. To mitigate this cost, numerous measurement-reduction strategies have been proposed, including commutativity-based term grouping\cite{verteletskyi_measurement_2020}, unitary partitioning\cite{izmaylov_unitary_2020,Measurement_reduction2021,Measurement_reduction2022}, classical shadow tomography tailored to chemistry~\cite{huang_predicting_2020,zhao_classicalshadow,hadfield_measurements_2022}, and low-rank factorizations\cite{motta_low_2021,huggins_efficient_2021}. Nevertheless, achieving chemical accuracy remains an open and critical challenge, particularly for variational approaches. These methods rely on iterative parameter optimization, which requires repeated gradient evaluations typically expressed as linear combinations of expectation values via the parameter-shift rule\cite{Evaluating_analytic_2019, kottmann_feasible_2021}. 

Here, we propose a hybrid approach that avoids iterative optimization of variational parameters shifting key update steps to classical preprocessing. Many previous developments in hybrid quantum-classical algorithms employ quantum circuits to prepare an (approximate) ground-state wave function, and can thus be interpreted within the Schrödinger picture, where the wave function evolves unitarily while the Hamiltonian remains fixed.
In contrast, this work adopts the Heisenberg picture, in which the Hamiltonian itself evolves. Specifically, we present a hybrid quantum-classical approach that seeks a quantum circuit capable of diagonalizing the Hamiltonian, using sequential Givens rotations\cite{Kivlichan_Givens_Rotations2018} tailored for electronic structure simulations. Although the Schrödinger and Heisenberg pictures are formally equivalent---the resulting unitary transformation ultimately yields the ground state when applied to a suitable initial state---the two lead to algorithmically distinct procedures; in particular, the adaptive gate update process proceeds in the reverse direction by transforming the Hamiltonian rather than the state. 
This perspective closely resembles that of the selective Projective Quantum Eigensolver (PQE) developed by Evangelista and co-workers\cite{EvangelistaPQE_2021,misiewicz_implementation_2024}, which employs a projection-based strategy to iteratively solve for the ground-state wave function without explicitly minimizing the energy. 
However, while PQE is designed for near-term, NISQ-friendly circuits, our proposal shifts as much of the computational burden as possible to the classical domain. It relies on sequential Hamiltonian transformations and deeper circuits, which are better suited to early FTQC than to NISQ devices, where coherence times and noise levels would make direct implementation challenging.

Whereas PQE solves projected residual equations within a fixed operator parameterization, its selective variants determine candidate operators by extracting residual information on quantum hardware based on real-time quantum evolution and measurements. In contrast, our method estimates candidate gates entirely through classical preprocessing, thereby reducing quantum workload associated with operator selection. Furthermore, by avoiding real-time evolution, our approach is naturally suited for applications involving non-Hermitian Hamiltonians. 

From the standpoint of circuit complexity, our framework aligns with recent resource-aware optimization strategies. Step-merging techniques in quantum imaginary-time evolution (QITE)~\cite{gomes_efficient_2020} reduce Trotter overhead by consolidating successive generators, while quaternion-based parameterizations of generalized controlled gates~\cite{Kurogi_2026} optimize free parameters to locally minimize implementation cost. In the same spirit, our protocol reduces circuit depth by merging repeated generators with small rotation angles, thereby eliminating redundant operations. However, rather than compressing time-evolution steps or optimizing a fixed gate structure, our simplifications arise from sequential Hamiltonian transformations coupled with perturbative generator merging. The optimization therefore acts at the level of operator selection and accumulation, shifting effort toward classical preprocessing and yielding systematically reorganized circuits compatible with early fault-tolerant architectures.

This paper is organized as follows.
Section~\ref{sec:theory} introduces the theoretical framework, including the Hamiltonian formalism, fermionic-to-qubit mappings, and efficient matrix-element selection.
We then present a truncation scheme and a cumulant-based expansion to enhance scalability, followed by a Monte Carlo sampling strategy and a circuit depth reduction procedure, and finally, provide an outline of the complete algorithm.
Section~\ref{sec:results} reports numerical results, beginning with noiseless benchmarks for N$_2$ at different geometries, including an analysis of residual amplitude distributions. We next analyze the H$_8$ system, where three geometric arrangements are compared using statistical diagnostics to quantify how the residual sparsity evolves across different molecular structures. We then perform a systematic benchmark on a strongly correlated H$_6$ chain to compare methods and to examine the trade-off between energy accuracy, measurement cost, and circuit depth. After having established algorithmic scaling and circuit-level costs under ideal conditions, we assess the impact of finite sampling noise on N$_2$ simulations under realistic shot budgets. 
Finally, Section.~\ref{sec:conclusion} summarizes our findings and discusses future directions for extending the method.

\section{Theory}
\label{sec:theory}
\subsection{General strategy of Jacobi method}
\label{sec:General strategy of Jacobi method}

We start with the Schr\"odinger equation,
\begin{align}
    \hat{H}\Psi = E\Psi
    \label{eq:Hamiltonian_eigenvalue} \\
    E = \langle \Psi|\hat H |\Psi\rangle \label{eq:Hamiltonian_expectation}
\end{align}
where the second-quantized Hamiltonian is given by 
\begin{align}
    \hat{H} &=\sum_{pq}^{} h^p_{q} \hat E^p_q + \sum_{pqrs}h^{pq}_{rs} \hat E^{pq}_{rs}\label{H1}
\end{align}
Here, $h^{p_1p_2\cdots p_n}_{q_1 q_2\cdots q_n}$ are $n$-body electron integrals associated with the normal-ordered $n$-body excitation operator
\begin{align}\hat E^{p_1p_2\cdots p_n}_{q_1 q_2\cdots q_n}
&= a^\dagger_{p_1} a^\dagger_{p_2} \cdots a^\dagger_{p_n} a_{q_n} \cdots a_{q_2} a_{q_1}
\label{excitation_term}
\end{align}
with $a^\dagger_{p}$ and $a_{p}$ being the creation and annihilation operators.  Below, we will occasionally use the shorthand $\hat E_i$ to indicate the $i$th combination of fermion operators.
The exact ground state wave function $|\Psi\rangle$ can be always written as
\begin{align}
    |\Psi\rangle = \hat U |\Phi_0\rangle
    \label{unitaryU}
\end{align}
where $\hat U$ is some unitary operator and $|\Phi_0\rangle$ is a reference determinant having nonzero overlap with the exact full configuration interaction (FCI) state; a typical choice is the Hartree-Fock (HF) determinant. In practice, algorithms such as VQE employ an "ans\"atze" that fixes the mathematical structure of $\hat U$, which is parametrized by variational parameters. 

One can instead write the Schr\"odinger equation as
\begin{align}
    \bar H |\Phi_0\rangle = E |\Phi_0\rangle\label{eq:1}
\end{align}
with $\bar H = \hat U^{\dagger} \hat H \hat U$ being the rotated Hamiltonian whose eigenstate is $|\Phi_0\rangle$. Therefore, it is regarded that the Hamiltonian is updated instead of the wave function. Projecting Eq.~(\ref{eq:1}) onto $|\Phi_0\rangle$ and other orthogonal determinants $|\Phi_\mu\rangle$  $(\mu\ne 0)$  yields the following set of equations:
\begin{align}
    E &= \langle \Phi_0| \bar H |\Phi_0\rangle\\
    0 &= \langle \Phi_\mu| \bar H |\Phi_0\rangle\label{eq:2}
\end{align}

\begin{figure*}
  \includegraphics[width=0.65\textwidth]{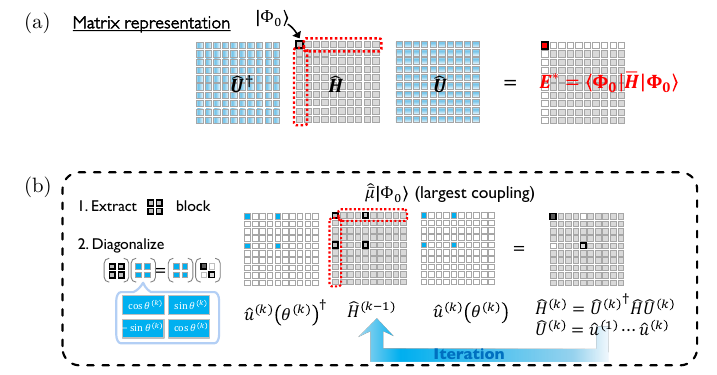}
  \caption{(a) General representation of the Hamiltonian transformation, 
  where the unitary operator eliminates off-diagonal couplings with respect to the Hartree--Fock determinant $|\Phi_0\rangle$ (red blocks).
  (b) Specific Hamiltonian update under deterministic selection, with {\bf G} highlighted in cyan boxes; white and gray entries denote zero and non-zero elements, respectively.}
  \label{fig:hamiltonian-both}
\end{figure*}

Fig.~\ref{fig:hamiltonian-both} illustrates the matrix representation of the transformation
\(\bar H = \hat U^\dagger \hat H \hat U\), in which the unitary operator zeroes out the off-diagonal couplings with respect to the Hartree--Fock determinant \(|\Phi_0\rangle\). 
Hence, our aim is to find $\hat U$ such that Eq.~(\ref{eq:2}) is satisfied. 
In practice, it is sufficient to find a unitary transformation that block-diagonalizes the Hamiltonian, although such a $\hat U$ remains highly nontrivial.

To this end, we employ a series of Givens rotations $\hat u$ to  build $\hat U$ as,
\begin{align}
    \hat U = \hat u^{(1)} \hat u^{(2)} \cdots
\end{align}

Each rotation acts in a two-dimensional subspace spanned by the reference determinant $|\Phi_0\rangle$ and a connected determinant $|\Phi_\mu\rangle$.
Therefore, we first introduce the Hermitian operator $\hat \mu$, called \textit{generator}, such that
\begin{align}
    |\Phi_\mu\rangle = \hat \mu |\Phi_0\rangle\label{eq:10}
\end{align}
and 
\begin{align}
|\Phi_0\rangle = \hat \mu |\Phi_{\mu}\rangle\label{eq:11}    
\end{align}
up to a phase factor. 
This dual relation, combined with the orthogonality condition $\langle\Phi_\mu|\Phi_0\rangle = \delta_{\mu,0}$, ensures that the operator
\begin{align}
    \hat u^{(k)} = e^{ \text{i} \theta^{(k)} \hat \mu^{(k)}}\label{eq:9}
\end{align}
performs a Givens rotation between the basis determinants $|\Phi_0\rangle$ and $|\Phi_{\mu}\rangle$ by an angle $\theta$.
Within this framework, the Hamiltonian is updated step-by-step: at the $k$th cycle, one obtains
\begin{align}
    \hat H^{(k)} = \hat u^{(k)\dagger}\hat H^{(k-1)} \hat u^{(k)}  
\end{align}

To find the appropriate operator $\hat \mu$ and the corresponding angle $\theta$ at each cycle, we follow the classical Jacobi algorithm in matrix diagonalization. Specifically, we compute the off-diagonal elements of the Hamiltonian with respect to $|\Phi_0\rangle$, and determine the determinant $|\Phi_{\bar \mu}\rangle$ that most interacts with $|\Phi_0\rangle$ via the present Hamiltonian $\hat H^{(k)}$. In other words, the determinant is selected by
\begin{align}
    \bar \mu = \arg \max_{\mu\ne 0} |\langle \Phi_\mu | \hat H^{(k)} |\Phi_0 \rangle|
    \label{determinant_max}
\end{align}
which serves as the next operator, i.e., $\hat \mu^{(k+1)} \leftarrow \hat {\bar \mu}$. 
To make this selection more transparent, it is convenient to introduce the following residual vector state
\begin{equation}
    |r^{(k)}\rangle = \left( \hat{H}^{(k)} - E^{(k)} \right) |\Phi_{0}\rangle
    = \sum_{\mu} c_\mu^{(k)} |\Phi_\mu\rangle,
    \label{eq:residual}
\end{equation}
where \( E^{(k)} = \langle \Phi_{0} | \hat{H}^{(k)} | \Phi_{0} \rangle \) is the energy expectation value in iteration \( k \), and \( c_\mu^{(k)} \in \mathbb{C} \) are the residual amplitudes. The selected determinant corresponds to the largest residual amplitude,
$c_{\bar \mu}^{(k)} = \langle\Phi_{\bar \mu}|\hat H^{(k)}|\Phi_0\rangle$. We perform the Givens rotation so that this dominant component is made zero at each cycle.

Despite its simplicity, this strategy effectively ensures a systematic reduction of the residual norm at each iteration as per the Jacobi method. In other words, 
\begin{align}
    \||r^{(k+1)}\rangle\| < \||r^{(k)}\rangle\|.
\end{align}
The convergence to the FCI limit is achieved when $\||r^{(k)}\rangle\| \rightarrow 0$ (see Eq.~(\ref{eq:2})).

Having determined the determinant with the largest coefficient in the residual, i.e., $|\Phi_{\bar \mu}\rangle = \hat {\bar \mu}|\Phi_0\rangle$, we now evaluate the rotation angle $\theta$.
We construct an effective two-dimensional Hamiltonian in the basis spanned by the reference Hartree--Fock state \( |\Phi_{0}\rangle \) and the selected determinant \( |\Phi_{\bar \mu}\rangle \):

\begin{align}
    {\bf M}^{(k)} &= \begin{pmatrix}
\langle \Phi_{0}|\hat H^{(k)}|\Phi_{0}\rangle & \langle \Phi_{0}|\hat H^{(k)}|\Phi_{\bar \mu}\rangle \\
\langle \Phi_{\bar \mu}|\hat H^{(k)}|\Phi_{0}\rangle & \langle \Phi_{\bar \mu}|\hat H^{(k)}|\Phi_{\bar \mu}\rangle
\end{pmatrix}
\nonumber\\
&= \begin{pmatrix}
E^{(k)} & c_{\bar \mu}^{(k)*} \\
c_{\bar \mu}^{(k)} & E^{(k)}_{\bar \mu}
\end{pmatrix}
\label{Matrix M}
\end{align}
where $E_{\bar \mu}^{(k)}$ corresponds to the energy expectation value of the chosen determinant $|\Phi_{\bar \mu}\rangle$ in the present Hamiltonian. These elements define the extracted block shown in Fig.~\ref{fig:hamiltonian-both}.

The proper Givens rotation in this subspace, 
\begin{align}
    {\bf G} = \begin{pmatrix}
        \cos\theta^{(k+1)} & \sin\theta^{(k+1)} \\
        -\sin\theta^{(k+1)} & 
        \cos\theta^{(k+1)}
    \end{pmatrix}
\end{align}
is defined so as to diagonalize this small effective Hamiltonian ${\bf M}^{(k)}$ 
\begin{align}
    {\bf G}^\dagger{\bf M}^{(k)} {\bf G}= \begin{pmatrix}
        E^{(k+1)} & 0\\
        0 & E^{(k)} + E^{(k)}_{\bar \mu}-E^{(k+1)}
        \label{eq:GMG}
        \end{pmatrix},
\end{align}
where we assume $E^{(k+1)}$ to be the lower eigenvalue, which reflects the minimum energy achievable within the current two-dimensional subspace. It is easy to verify that
\begin{align}
    E^{(k+1)} = \frac{E^{(k)}+E^{(k)}_{\bar \mu}}{2} - \sqrt{\left(\frac{E^{(k)}-E^{(k)}_{\bar \mu}}{2}\right)^2 +  |c_{\bar \mu}^{(k)}|^2}.
\end{align}

The optimal rotation angle \(  \theta^{ (k+1)} \) can be computed analytically from the off-diagonal coupling and energy difference:
\begin{equation}
\theta^{(k+1)} =  \frac{1}{2} \tan^{-1} \left( \frac{2 c_{\bar \mu}^{(k)}}{E^{(k)} - E^{(k)}_{\bar \mu}} \right),
\label{theta_formula}
\end{equation}
where we have assumed that the transformed Hamiltonian and $c_{\bar \mu}^{(k)}$ are real in the basis of $\{|\Phi_\mu\rangle\}$ for simplicity, but it is straightforward to extend this result to complex Hermitian Hamiltonians.
With a similar treatment, our scheme can be readily extended to non-Hermitian Hamiltonians, although such an extension is beyond the focus of the present study.

For the aforementioned scheme to work out, an operator basis for $\hat \mu$ should be chosen that possesses the transfer character of Eqs.~(\ref{eq:10})--(\ref{eq:11}). Below, we consider two different approaches for such an operator basis.

\subsection{Pauli Operator}
\label{sec:Pauli}

To simulate fermionic systems on quantum hardware, it is necessary to map second-quantized fermionic operators to qubit operators. In this work, we adopt the Jordan--Wigner (JW) transformation\cite{article:JordanWigner1928}, a widely used scheme that maps a system of \( N \) fermionic modes onto \( N \) qubits, assigning one qubit per mode, while exactly preserving the canonical anticommutation relations. 
After applying the JW transformation, the electronic Hamiltonian, initially expressed in terms of creation and annihilation operators, is represented as a weighted sum of $N_P$ tensor products of Pauli operators acting on qubits:
\begin{align}
    \hat H = \sum_i^{N_P} h_i \hat P_i \label{H_Pauli}
\end{align} 
where each term in the resulting qubit Hamiltonian involves combinations of the identity \( I \) and the Pauli matrices \( X \), \( Y \), and \( Z \),
\begin{align}
    \hat P \in \{I,X,Y,Z\}^{\otimes n_{\rm qubits}}
\end{align}
and $h_i$ is the coefficient associated with $\hat P_i$. This structure holds for $\hat H^{(k)}$ and we indicate it by adding the superscript $(k)$.

In the Pauli-based approach, we simply choose a single Pauli operator $\hat P_\mu$ for the generator basis $\hat \mu$.
Owing to the self-inverse property of Pauli strings \( (\hat{P}_\mu^2 = I) \), the relations Eqs.~(\ref{eq:10}-\ref{eq:11}) hold and the operator \( e^{i\theta \hat{P}_\mu} \) admits a closed-form expansion via the BCH formula. Specifically, each individual term $\hat P_i$ from the decomposed Hamiltonian undergoes the transformation, if $\left[\hat P_i, \hat P_\mu\right] \ne 0$:

\begin{align}
     e^{-\text{i}\theta\hat{P}_\mu} \hat{P}_i
     e^{\text{i}\theta\hat{P}_\mu} 
     &= 
    (\cos^2\theta - \sin^2\theta)\hat P_i 
    + \tfrac{\text{i}}{2}\sin(2\theta)[\hat P_i, \hat P_\mu]
      \label{BCHPauli}
\end{align}
It is obvious that if $\left[\hat P_i, \hat P_\mu\right] = 0$, the Hamiltonian term remains unchanged under the transformation.

In constructing the excitation operators \( \hat{P}_\mu \) that act on the Hartree--Fock reference state to generate new configurations \( |\Phi_\mu\rangle = \hat{P}_\mu |\Phi_{0}\rangle \), we adopt a minimal and symmetry-respecting approach. Specifically, we employ Pauli strings composed of \( X \) operators acting on the qubits whose occupation changes between \( |\Phi_{0}\rangle \) and \( |\Phi_\mu\rangle \), with a single \( Y \) operator replacing one of the \( X \). 
This construction guarantees that the resulting unitary \( e^{i\theta \hat{P}_\mu} \) is real, thereby preserving the real-valuedness of the transformed Hamiltonian. 
Although more general Pauli strings that involve arbitrary combinations of \( X \), \( Y \), and \( Z \) operators could be employed, we restrict ourselves to this minimal set to reduce circuit complexity and maintain compatibility with real-valued simulations. However, one important drawback of the Pauli-based approach is that it breaks many symmetries in the transformed Hamiltonian. As has been discussed,\cite{Tsuchimochi_adapt_2022, bertels_symmetry_2022} such symmetry violation can lead to relatively slow convergence.

\subsection{Fermionic Operator}
\label{sec:Fermionic Operator}

An alternative approach is based on the excitation and de-excitation anti-Hermitian operators $\hat{A}^{(k)}$ such that:
\begin{equation}
        \hat{\mu} = -\text{i}\hat{A} = -\text{i}(\hat{E}-\hat{E}^{\dag})
        \label{Fermion exp}
\end{equation}
where $\hat E$ is an excitation operator as defined in Eq.~({\ref{excitation_term}}).

The fermionic Givens rotation
\begin{align}
    \hat u = e^{\theta \hat A}
\end{align}
yields a transformed Hamiltonian that preserves both particle-number and $S_z$ symmetries of the original Hamiltonian. Following the formalism introduced by Evangelista and Magoulas on exact closed-form unitary transformations of single fermionic operators $\{\hat E_i\}$\cite{evangelista2024exactclosedformunitarytransformations}, we exploit the nilpotency condition of the operator \( \hat{A} \), i.e. \(\hat{A}^3 = -\hat{A}\), to carry out the BCH expansion on each Hamiltonian term individually. 

If an excitation operator $\hat E_i$ commutes with the anti-Hermitian operator $\hat A$, the rotated term naturally remains unchanged. Otherwise, two distinct closed forms arise depending on the specific commutation relation. In particular, if $\hat{A} \left[\hat{E}_i, \hat{A} \right]\hat{A}=0$,  then
\begin{align}
    e^{-\theta \hat A} \hat E_i e^{\theta \hat A} = \hat{E}_i 
    &+ \sin\theta\left[\hat{E}_i, \hat{A}\right] + \left(1 - \cos{\theta}\right) \left[\left[\hat{E}_i, \hat{A}\right], \hat{A}\right]
    \label{eq:order1}
\end{align}
whereas in the general case,
\vspace{0.5cm}
\begin{align}
  e^{-\theta \hat A} \hat E_i e^{\theta \hat A} = \hat{E}_i
    &+ \frac{1}{2} \sin2\theta \left[\hat{E}_i, \hat{A}\right] + \frac{1}{2} \sin^2\theta \left[\left[\hat{E}_i, \hat{A}\right], \hat{A}\right]
    \label{eq:order2}
\end{align}
These expressions provide an exact transformation of each Hamiltonian term under the unitary rotation.\cite{evangelista2024exactclosedformunitarytransformations} 
Due to the nested structure of commutators in the above expressions, evaluating the updated Hamiltonian becomes increasingly complex, as higher-rank operators may be generated. Notably, if \( \hat{A} \) corresponds to an excitation of rank greater than one, the resulting transformation can introduce terms of even higher excitation rank.
Specifically, performing the single and double commutator terms, their excitation ranks become:
\begin{align}
    {\rm rank}\left(\left[
    {\hat E_i},\hat A\right]\right) & = {\rm rank}(
    {\hat E_i}) + {\rm rank}({\hat A})-1\label{rankcommutator}\\
  {\rm rank}\left(\left[\left[
    {\hat E_i},\hat A\right],\hat A\right]\right)&= {\rm rank}(
    {\hat E_i}) + 2\;{\rm rank}({\hat A})-2\label{rankdoublecommutator}
\end{align}

Similarly to the Pauli approach discussed in Section \ref{sec:Pauli}, we take the simplest and unique approach to construct $\hat A_\mu$ in our algorithm. 
It should be noted that a general excitation operator of rank $n$ is expressed as

\begin{align}
    \hat E^{p_1\cdots p_m r_1\cdots r_{l}}_{q_1\cdots q_m r_1\cdots r_{l}}
    =     \hat E^{p_1\cdots p_m}_{q_1\cdots q_m}
    \hat E_{r_1\cdots r_l}^{r_1\cdots r_l}
    \label{eq:pure_and_spectator}
\end{align}
where $n=m+l$.  Here, the index sets $\{p_1, \cdots, p_m\}$, $\{q_1, \cdots, q_m\}$, and $\{r_1, \cdots, r_l\}$ are pairwise disjoint, i.e., they share no common elements. 
The last set of indices appears both in the creation and annihilation parts, and is referred to as a \textit{spectator index}, corresponding to the operator pair \( a_r^\dagger a_r \), which acts trivially on $|\Phi_0\rangle$ and yields no net excitation. In contrast, the first two index sets generate genuine excitations (unless they annihilate $|\Phi_0\rangle$) and are therefore called ``pure'' indices.
To reduce complexity, $\hat A$ is selected so that it only contains pure excitation. This distinction between pure and spectator indices will become essential in the cumulant decomposition presented in Sec.~\ref{sec:Cumulant Decomposition}.

\subsection{Quantum Jacobi}
\label{sec:Quantum Jacobi}

Although the preceding discussion is exact and the quantum Jacobi procedure formally converges to the FCI solution, performing the BCH expansion exactly on classical computers is infeasible, as the number of Hamiltonian terms grows rapidly with increasing $k$.
To render the algorithm tractable, we instead leverage quantum computers. Specifically, we now turn to the Schr\"odinger picture, where the wave function is evolved through a sequence of Givens rotations:
\begin{align}
\hat U^{(k)}|\Phi_0\rangle = \hat u^{(1)} \hat u^{(2)}\cdots \hat u^{(k)}|\Phi_0\rangle
\end{align}
This allows for the determination of $E^{(k)}$ for instance, by measuring the expectation value,
\begin{align}
    E^{(k)} = \langle \Phi_0|\hat U^{(k)\dagger} \hat H \hat U^{(k)}|\Phi_0\rangle \label{eq:E^(k)}
\end{align}
In practice, however, measuring $E^{(k)}$ is not needed because it corresponds to the eigenvalue of the previous \( \mathbf{M}^{(k-1)} \), which can be classically obtained once the required matrix elements are known. Therefore, the essential role of quantum computers in the proposed quantum Jacobi (QJ) algorithm is the evaluation of these matrix elements.  
Once the appropriate generator $\hat {\bar\mu} = \hat \mu^{(k+1)}$ is identified for the next update, the remaining matrix elements in Eq.~(\ref{Matrix M}), namely, $c_{\bar \mu}^{(k)}$ and $E^{(k)}_{\bar \mu}$, can be evaluated in a quantum-classical hybrid manner:
\begin{align}
E^{(k)}_{\bar \mu} &= \langle \Phi_0|e^{-\text{i}\frac{\pi}{2}\hat {\bar  \mu}} \hat U^{(k)\dagger} \hat H \hat U^{(k)} e^{\text{i}\frac{\pi}{2}\hat {\bar  \mu}}   |\Phi_0\rangle\label{eq:Ekmu}\\
c_{\bar \mu}^{(k)} &= \langle \Phi_0|e^{-\text{i}\frac{\pi}{4}\hat{\bar  \mu}} \hat U^{(k)\dagger} \hat H \hat U^{(k)} e^{\text{i}\frac{\pi}{4}\hat {\bar  \mu}} |\Phi_0\rangle - \frac{E^{(k)}+E^{(k)}_{\bar\mu}}{2}\label{eq:cmu}
\end{align}
Therefore, forming the effective $2\times2$ Hamiltonian ${\bf M}^{(k)}$ requires only the two expectation values in Eqs.~(\ref{eq:Ekmu}) and (\ref{eq:cmu}). The resulting matrix ${\bf M}^{(k)}$ is then diagonalized {\it classically} to obtain the updated energy \( E^{(k+1)} \) and the optimal Givens rotation angle \( \theta^{(k+1)} \) using Eq.~(\ref{theta_formula}). This hybrid workflow ensures that quantum resources are used efficiently, restricted only to measuring quantities that cannot be inferred from the classical update, and that the Hamiltonian transformation and rotation angle optimization remain entirely classical. Consequently, the remaining task is to identify an appropriate generator $\hat {\bar \mu}$ according to the residual vector $|r^{(k)}\rangle$ (Eq.~(\ref{eq:residual})).

One possible scheme for estimating the residual vector \( |r^{(k)}\rangle \) is to apply a short-time evolution operator on quantum hardware, similar to the strategy used in the SPQE approach introduced by Evangelista \textit{et al.}~\cite{EvangelistaPQE_2021}. 
This leads to the following quantum state:
\begin{align}
    |R^{(k)}\rangle &=\hat{U}^{(k)\dagger} e^{-\text{i} \hat{H} \Delta t} \hat{U}^{(k)} |\Phi_0\rangle \nonumber \\
    &= |\Phi_0\rangle - \text{i} \Delta t\, \hat{U}^{(k)\dagger} \hat{H} \hat{U}^{(k)} |\Phi_0\rangle + \mathcal{O}(\Delta t^2),
    \label{Residual_SPQE}
\end{align}
where \( \Delta t \) is a small evolution time. 
Eq.~(\ref{Residual_SPQE}) can be regarded as an approximation to $|r^{(k)}\rangle$ once the $|\Phi_0\rangle$ component is removed. Consequently, sampling bit strings from $|R^{(k)}\rangle$ by quantum measurements then reveals dominant configurations, with probabilities proportional to squared amplitudes.  
While conceptually appealing and naturally suited to quantum hardware, this approach introduces two major challenges: (a) the measurement cost increases due to the need for accurate resolution of small residual amplitudes, and (b) the method relies on Hermiticity of the Hamiltonian, making it nontrivial to extend to non-Hermitian or complex-symmetric formulations, which may appear in more general settings.

Given these challenges associated with real-time quantum evolution, we adopt a strategy in which the Hamiltonian is classically updated using an {\it approximate} BCH expansion $\bar H_{\rm approx}^{(k)}$ just to estimate ${\bar \mu}$, while the relevant matrix elements \( E^{(k)}_{\bar \mu} \) and \( c_{\bar \mu}^{(k)} \) are still estimated on quantum hardware. In other words, generator selection is performed by computing the approximate residual vector
\begin{align}
    |\tilde r^{(k)}\rangle = (\bar H_{\rm approx}^{(k)} -E^{(k)}_{\rm approx})|\Phi_0\rangle =\sum_{\mu} \tilde c_\mu^{(k)} |\Phi_\mu\rangle \label{eq:approximate_r}
\end{align}
which can be done efficiently with classical computers: the computational cost scales linearly with the number of terms in $\bar H_{\rm approx}^{(k)}$.
Below, we propose several schemes that allow us to maintain a manageable number of Hamiltonian terms without exponential growth in operator complexity while minimizing the loss of accuracy in the energy estimation.

\subsection{Truncation and Approximation}
\label{sec:Truncation and Approximation}

We introduce a structured truncation protocol to control the number of terms generated in both Pauli and Fermionic representations. Additionally, we present a cumulant-based decomposition scheme tailored for the Fermionic framework, designed to reduce the complexity of high-rank excitation operators and simplify the Hamiltonian representation.

\subsubsection{Hamiltonian Truncation}
\label{sec:Amplitude and Rank-Driven Operator Truncation}

It is evident that the importance of each transformed Hamiltonian term for maintaining the similarity between the exact $|r^{(k)}\rangle$ and the approximate $|\tilde r^{(k)}\rangle$ is related to the magnitude of its amplitude. 
Motivated by this observation, the simplest truncation scheme is to discard all terms with \( |h_i| < \epsilon \) at each cycle $k$, where \( \epsilon \) is a predefined amplitude threshold. 
While straightforward and easily applicable to both Pauli- and fermionic-operator based quantum Jacobi, this approach risks eliminating important higher-order excitations that are essential for describing strong correlation. Conversely, some terms with large coefficients may contribute little or nothing to $|\tilde r^{(k)}\rangle$.  

To address this, we introduce a rank-aware truncation scheme tailored to the fermion operator basis, by taking advantage of the concept of excitation rank $n$.  
Specifically, we retain all terms with \( n \leq 2 \) unconditionally, as they represent physically dominant one- and two-body processes. Higher-rank terms (\( n > 2 \)) are retained only if their coefficients satisfy \( |h_i| \geq \epsilon \). In addition, operator terms with excitation rank exceeding the number of electrons are omitted entirely, as they annihilate the HF determinant. This truncation is generally justified, since such operators will never reduce their ranks in subsequent BCH expansions as shown in Eqs.~(\ref{rankcommutator}-\ref{rankdoublecommutator}). This rank-aware truncation exploits the structure of the BCH expansion, where nested commutators generate high-rank terms with rapidly decaying amplitudes.

\subsubsection{Cumulant Decomposition}
\label{sec:Cumulant Decomposition}

Independently of the truncation scheme described above, we consider a cumulant-based decomposition of high-order operators. Cumulant expansions provide a systematic framework for expressing many-body quantities in terms of lower-order correlations, enabling controlled approximations of complex quantum observables. In particular, the decomposition of the \( p \)-body reduced density matrix (RDM) into products of lower-order terms has been extensively investigated within reduced density matrix theory~\cite{PhysRevA.47.979,https://doi.org/10.1002/qua.560510605,PhysRevLett.76.1039,PhysRevA.56.2648,PhysRevA.57.4219,article:Approximatesolutionforelectroncorrelation}.
Mazziotti introduced a binomial expansion of the \( p \)-body hole density matrix using an antisymmetrized (Grassmann) tensor product between RDMs and identity operators~\cite{PhysRevA.57.4219}.
Truncating the cumulant hierarchy at a given rank within this expansion offers a controlled approximation strategy, often reducing computational cost while maintaining accuracy.
More recently, Yanai and Chan~\cite{article:CanonicaltranformationMR2006} demonstrated a decomposition of three-body excitation operators in terms of one- and two-body components:
\begin{equation}
\begin{aligned}
    a^{\dagger}_{i} a^{\dagger}_{j} a^{\dagger}_{k} a_{l} a_{m} a_{n}  
    &= 9\langle a^{\dagger}_{i} a_{n} \rangle \wedge (a^{\dagger}_{j} a^{\dagger}_{k} a_{l} a_{m}) \\
    &\quad - 12\langle a^{\dagger}_{i} a_{n} \rangle \wedge \langle a^{\dagger}_{j} a_{m} \rangle \wedge (a^{\dagger}_{k} a_{l}),
\end{aligned}
\label{Yanai_Chan_3bodyRDM}
\end{equation}
where the contractions \(\langle a^{\dagger}_{p} a_{q} \rangle\) are evaluated with respect to the reference wave function, and the wedge product enforces the appropriate fermionic antisymmetry with a prefactor of $1/(p!)^2$. This decomposition facilitates the approximate representation of high-rank excitation operators using lower-order building blocks, consistent with cumulant-based truncation strategies.

Inspired by these works, our aim here is to apply the cumulant-decomposition to high-rank operators, specifically three-body or more, to approximate the canonical BCH expansion. The chief difference from the original proposal of Yanai and Chan lies in the fact that our reference wave function is the HF determinant $|\Phi_0\rangle$, which makes the decomposition particularly simple, since the contraction is the Kronecker's delta itself. 
However, a key caveat in applying the full cumulant decomposition to our case is that it is too simple; most contractions vanish, potentially eliminating important contributions. Moreover, the resulting effective Hamiltonian  $\bar H_{\rm approx}$ contains only one- and two-body operators, and thus neglects important higher excitations. Namely, the classically approximated residual vector $|\tilde r^{(k)}\rangle$ would be expanded only by singly and doubly excited determinants. This may be acceptable for weakly correlated systems where UCCSD is an accurate workhorse, but such high-rank excitations are deemed to be essential for strongly correlated systems and thus should not be cumulant-decomposed. To avoid these issues while effectively incorporating the idea of cumulant decomposition, we design a scheme specifically tailored to the HF reference.

At each iteration $k$, high-rank ($n>2$) excitation operators $\hat E_i$ are first screened by their coefficients $h_i$. Those with $|h_i| > \kappa$ ($\kappa > \epsilon$ is another threshold) are kept, since they are likely to remain significant in the subsequent BCH expansion.
Other high-rank operators with \( |h_i| < \kappa \) are treated using a \textit{restricted} cumulant decomposition, in which the pure excitation part \( \hat E_{q_1\cdots q_m}^{p_1\cdots p_m} \) (defined in Eq.~\ref{eq:pure_and_spectator}) is left unchanged, and contractions are applied only to the spectator component \( \hat E_{r_1\cdots r_l}^{r_1\cdots r_l} \). 
Noting that contraction with unoccupied indices gives zero for the HF state, we further simplify the scheme by discarding operators that include unoccupied pairs in $\hat E_{r_1\cdots r_l}^{r_1\cdots r_l}$. We then repeatedly decompose the occupied spectator operator with the appropriate normalization factors:
\begin{align}
\hat{E}^{r_1 \cdots r_l}_{r_1 \cdots r_l} 
&\;\longrightarrow\; \frac{1}{l} \hat{E}^{r_1 \cdots r_{l-1}}_{r_1 \cdots r_{l-1}} \wedge \langle \hat{E}^{r_l}_{r_l} \rangle, \nonumber \\
&\;\longrightarrow\; \frac{2}{l(l-1)} \hat{E}^{r_1 \cdots r_{l-2}}_{r_1 \cdots r_{l-2}} \wedge \langle \hat{E}^{r_{l-1}}_{r_{l-1}} \rangle \wedge \langle \hat{E}^{r_l}_{r_l} \rangle, \nonumber \\
&\;\longrightarrow\; \cdots \label{eq:39}
\end{align}
This recursive contraction proceeds until the total operator (including the pure excitation part) reduces to a two-body form, or terminates earlier if all spectator pairs have been fully contracted. 
Note that the wedge product in Eq.(\ref{eq:39}) carries no prefactor; it acts solely as an anti-symmetrizer.
In practice, this yields three distinct cases depending on the number of pure excitation indices $m$: 
\begin{enumerate}
\item \textbf{Full spectator operators} ($m=0$): 
All orbitals are spectators, and the operator reduces to a linear combination of two-body spectator operators through recursive contractions of spectator pairs.

\item \textbf{Single pure excitation} ($m=1$): 
The operator again reduces to a linear combination of two-body operators, each containing one spectator pair and one pure-excitation pair. The overall amplitude is scaled by $1/l$ to distribute it uniformly across all spectator choices.

\item \textbf{Multiple pure excitations} ($m>1$): 
Spectator contractions are fully applied individually until the operator reduces to a pure excitation form. This means that $\hat{E}^{r_1 \cdots r_l}_{r_1 \cdots r_l}$ is replaced by 1.
\end{enumerate}

\subsection{Monte Carlo Sampling}
\label{sec:Monte-Carlo sampling}

While the deterministic selection strategy that targets the largest (approximate) residual amplitude in $|\tilde r^{(k)}\rangle$ is effective at reducing the residual norm at each iteration, it eventually leads to stagnation. In particular, when the same generator is repeatedly selected across successive iterations due to the approximation in $\hat H_{\rm approx}^{(k)}$, i.e., \( {\bar{\mu}} = \mu^{(k)}\), there is no further effective update on the quantum side, since \begin{align}
    e^{-\text{i}\theta^{(k)} \hat \mu^{(k)}} e^{-\text{i}\theta_{\bar \mu} \hat {\bar \mu}} = e^{-\text{i}(\theta^{(k)} + \theta_{\bar \mu}) \hat \mu^{(k)}}
\end{align} where $\theta^{(k)}$ is already optimal and therefore $\theta_{\bar \mu} = 0$. On the classical side, this in turn means that there is virtually no BCH expansion, resulting in $\hat H_{\rm approx}^{(k+1)} = \hat H_{\rm approx}^{(k)}$, and the algorithm becomes stuck.

To address this limitation and promote broader exploration of determinant space, we introduce a complementary stochastic selection strategy inspired by the measurement-based algorithm based on the short-time evolution described above. Specifically, we resort to Monte Carlo sampling~\cite{greer_estimating_1995}. Instead of always selecting the excitation with the largest amplitude, candidate excitations are sampled probabilistically based on their residual weights.
This approach is considered particularly suitable in situations where the approximate and exact Hamiltonians are similar overall, but the approximate residual vector includes a few outlier configurations that do not appear in the exact one.
We define a reduced configuration space \( \mathcal{A}_\text{MC}^{(k)} \) that excludes the previously selected excitation \( \mu^{(k)} \) so that it will not be considered in the sampling.
The residual amplitudes \( \{ \tilde c_\mu^{(k)} \}_{\mu \in \mathcal{A}_\text{MC}^{(k)}} \) are normalized as
\begin{equation}
\bar {c}_\mu^{(k)} = 
\frac{\tilde c_\mu^{(k)}}{\sqrt{ \displaystyle\sum_{\mu \in \mathcal{A}_\text{MC}^{(k)}} |\tilde c_\mu^{(k)}|^2 }},
\label{eq:c_tilde}
\end{equation}
yielding the probability distribution
\begin{equation}
\bar p_\mu = |\bar {c}_\mu^{(k)}|^2
\label{eq:proba_dist}
\end{equation}
based on which the next generator is selected randomly. This procedure can be viewed as a virtual ``measurement'' of the pseudo-quantum state
\begin{align}
    |\bar r^{(k)}\rangle = \sum_{\mu} \bar c_\mu^{(k)} |\Phi_\mu\rangle
\end{align}
which is appropriately normalized.

This simple yet principled sampling scheme ensures that large-amplitude excitations are favored, while still allowing for non-negligible probability of selecting lower-amplitude excitations, thereby facilitating broader traversal of the excitation manifold.
While more sophisticated strategies could further enhance sampling diversity, we find that this basic stochastic formulation already offers a practical trade-off between exploration and exploitation within the Jacobi framework.

\begin{figure*}
  \includegraphics[width=\textwidth]{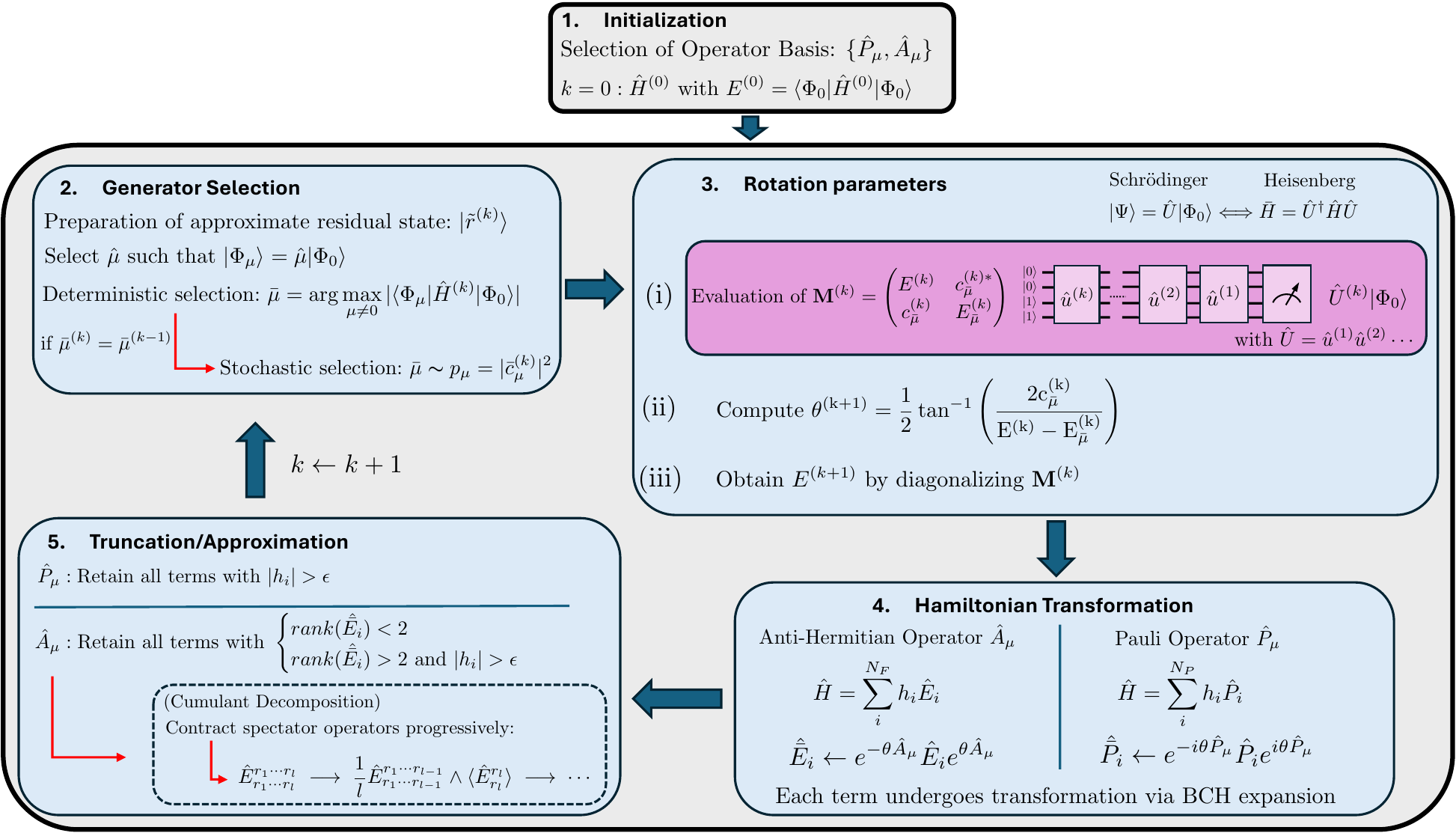}
  \caption{Schematic outline of the algorithmic workflow. Blue blocks indicate classical processing steps, while pink block represents operations in quantum circuits.}
  \label{Outline}
\end{figure*}

\begin{figure*}[t]  
  \centering
  \includegraphics[width=\textwidth]{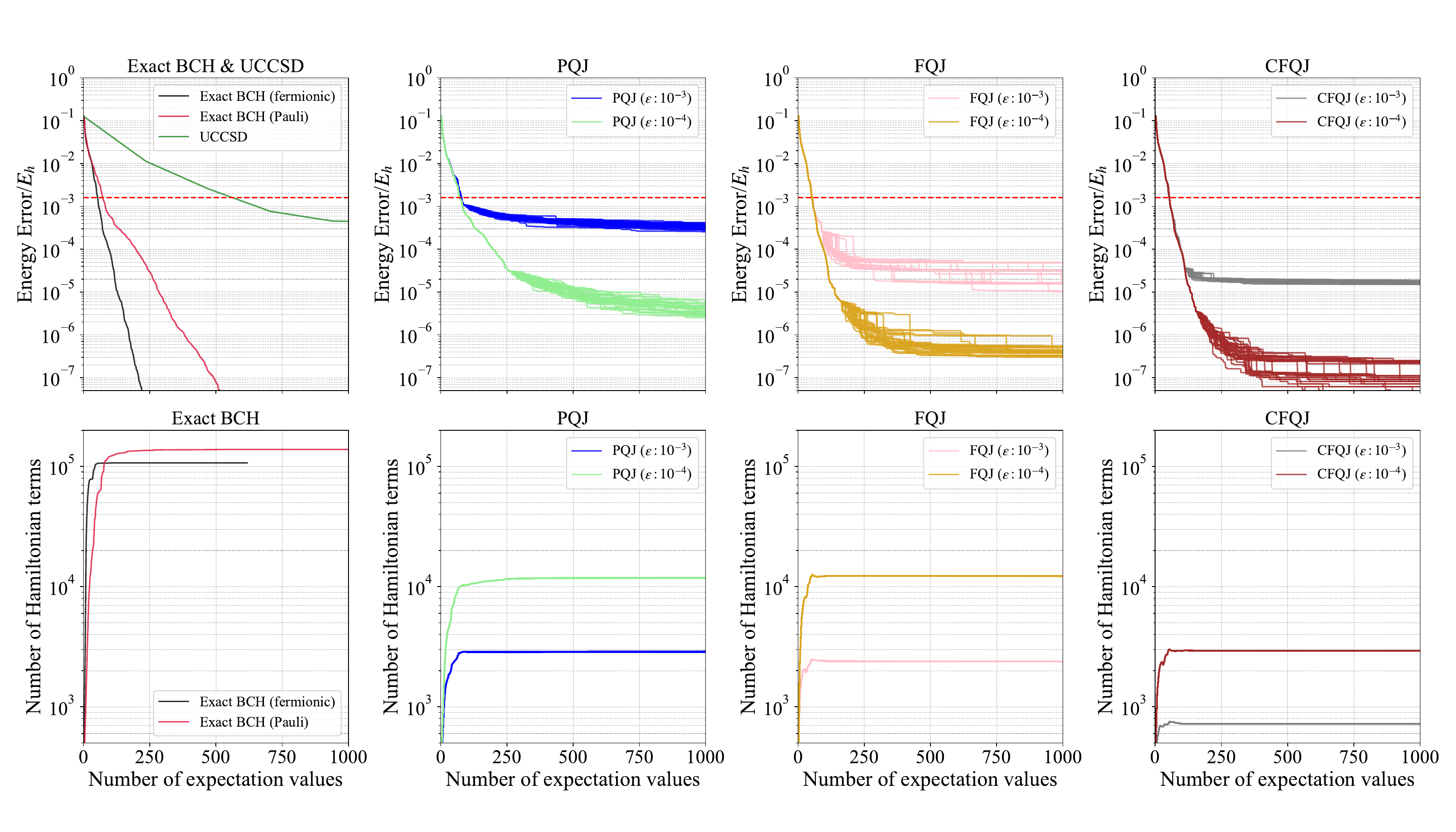}
  \caption{Energy error relative to FCI (top panels) and Hamiltonian term count (bottom panels) as a function of the number of expectation values for N$_2$ at the near-equilibrium bond length of 1.0977~\AA. Results are shown for PQJ, FQJ, and CFQJ at truncation thresholds $\epsilon = 10^{-3}$ and $10^{-4}$, with comparisons to the exact BCH expansion and UCCSD. The red dashed lines in the top panels denote the chemical accuracy threshold.}
  \label{N2Eerror_1.0977}
\end{figure*}

\subsection{Gate reduction by angle merging}
\label{sec:CNOT}

During the iterative operator-selection procedure, each cycle $k$ generates an operator $\hat{\mu}^{(k)}$ and the corresponding rotation angle $\theta^{(k)}$ based on it. A direct implementation of the resulting sequence of parametrized unitaries may therefore contain repeated occurrences of the same operator with different rotation angles, making the circuit deeper and introducing an unnecessarily large number of redundant parameters. Reducing this complexity is thus essential for practical computations even in the context of FTQC.

In general, for any unitary sequence containing two occurrences of the same generator, i.e., $\hat \mu^{(k')} \equiv\hat \mu^{(k)}$ ($k'<k$), one obtains
\begin{align}
	\hat U^{(k)} &=  \cdots e^{-\text{i} \theta^{(k')}\hat \mu^{(k')}} \cdots  e^{-\text{i} \theta^{(k)} \hat \mu^{(k)}}\nonumber\\
	& =\cdots e^{-\text{i} (\theta^{(k')}+\theta^{(k)})\hat \mu^{(k')}} \cdots 
	+ O\left((\theta^{(k)})^2\right)
\end{align} 
where the last term arises from the non-commutativity between $\hat \mu^{(k)}$ and all other operators appearing between the two occurrences.
This relation shows that, when $\theta^{(k)}$ is sufficiently small, the two repeated gates can be merged in to a single gate by updating the parameter as 
\begin{align}
	\theta^{(k')} \leftarrow \theta^{(k')}+\theta^{(k)} \label{eq:merge}
\end{align} 
The resulting circuit provides a first-order approximation to the original unitary $\hat U^{(k)}$ while reducing circuit depth. This idea is closely related to strategies previously explored in imaginary-time evolution to mitigate Trotterization costs\cite{gomes_efficient_2020}.

For consistency, the classical approximation $\bar H^{(k)}_{\rm approx}$ should then be reconstructed using the updated parameter $\theta^{(k')}$ for subsequent operator selection. In practice, however, this inconsistency has a negligible impact on our results, particularly when combined with Monte Carlo sampling, while still enabling a substantial reduction in circuit depth, as shown later.

\subsection{Outline of the Procedure}
\label{sec:Outline of the Procedure}

Here we present an overview of the iterative procedure. The algorithm switches between classical and quantum computation to select new operators, evaluate matrix elements and perform the transformation of the $k^{\text{th}}$ Hamiltonian.
The main components are summarized as follows.

\begin{enumerate}
    \item \textit{Initialization:} Set the initial Hamiltonian \( \hat{H}^{(0)} \), with classical energy estimate \( E^{(0)} = \langle \Phi_0 | \hat{H}^{(0)} | \Phi_0 \rangle \), which is the HF energy.

    \item \textit{Classical selection of the generator:} At each iteration \( k \), we \textit{classically} evaluate the approximate residual state \( |\tilde{r}^{(k)}\rangle \) to identify the next generator \( \bar{\mu} \). This step is initially performed deterministically by selecting the excitation with the largest residual amplitude, i.e., \( \bar{\mu} \) (see Eq.~(\ref{determinant_max})). Alternatively, in the Monte Carlo variant, the residual amplitudes are first normalized as \( \tilde{c}_\mu^{(k)} \), yielding a probability distribution \( \bar p_\mu \), from which the excitation operator is sampled stochastically (see Eqs.~(\ref{eq:c_tilde}) and (\ref{eq:proba_dist})).

    \item \textit{Rotation Parameters}: After having selected the generator $\bar \mu$, we construct the effective \( 2 \times 2 \) Hamiltonian matrix ${\bf M}^{(k)}$ in the basis of the reference state \( |\Phi_{0}\rangle \) and the target excited determinant $|\Phi_{\bar \mu}\rangle$, as defined in Eq.~(\ref{Matrix M}). The matrix elements are measured on a quantum device according to Eqs.~(\ref{eq:Ekmu}) and (\ref{eq:cmu}). The rotation angle \( \theta^{(k)} \) corresponding to the selected excitation is derived from the off-diagonal coupling and energy difference according to Eq.~(\ref{theta_formula}).

    \item \textit{Hamiltonian Transformation}:
    The selected operator and corresponding parameter \( \theta^{(k)} \) are applied via the exact closed-form BCH expansion (see Eq.~(\ref{BCHPauli}) for the Pauli-based scheme and Eqs.~(\ref{eq:order1}) and (\ref{eq:order2}) for the fermionic scheme).

    \item \textit{Truncation and Approximation}:
    To control the growth of operator number, the transformed Hamiltonian 
    is truncated on the basis of operator rank and/or amplitude of the Hamiltonian terms. For the fermionic scheme, the Hamiltonian can also be approximated using cumulant decomposition.
\end{enumerate}

\section{Results and Discussion}
\label{sec:results}

All numerical simulations were performed using {\sc Quket}, except for SPQE, for which we have used {\sc QForte}\cite{stair_qforte_2022}. {\sc Quket} is a Python-based quantum chemistry package\cite{QUKET}, {\sc Quket} integrating with {\sc PySCF}\cite{pyscf_2017,sun_recent_2020} for generating molecular orbitals and electronic integrals, {\sc OpenFermion}\cite{mcclean2019openfermionelectronicstructurepackage} for mapping fermionic operators to qubit operators, and {\sc Qulacs}\cite{Suzuki_2021} for quantum simulations.
The minimal STO-6G basis is used throughout this work to focus on the behavior of the methods within fixed active spaces, without introducing additional complexity from larger basis sets.

For the truncated QJ approaches, we initially employ deterministic operator selection and switch to a stochastic protocol once the same operator is selected in consecutive steps, as described above.

\begin{figure*}[t]  
  \includegraphics[width=\textwidth]{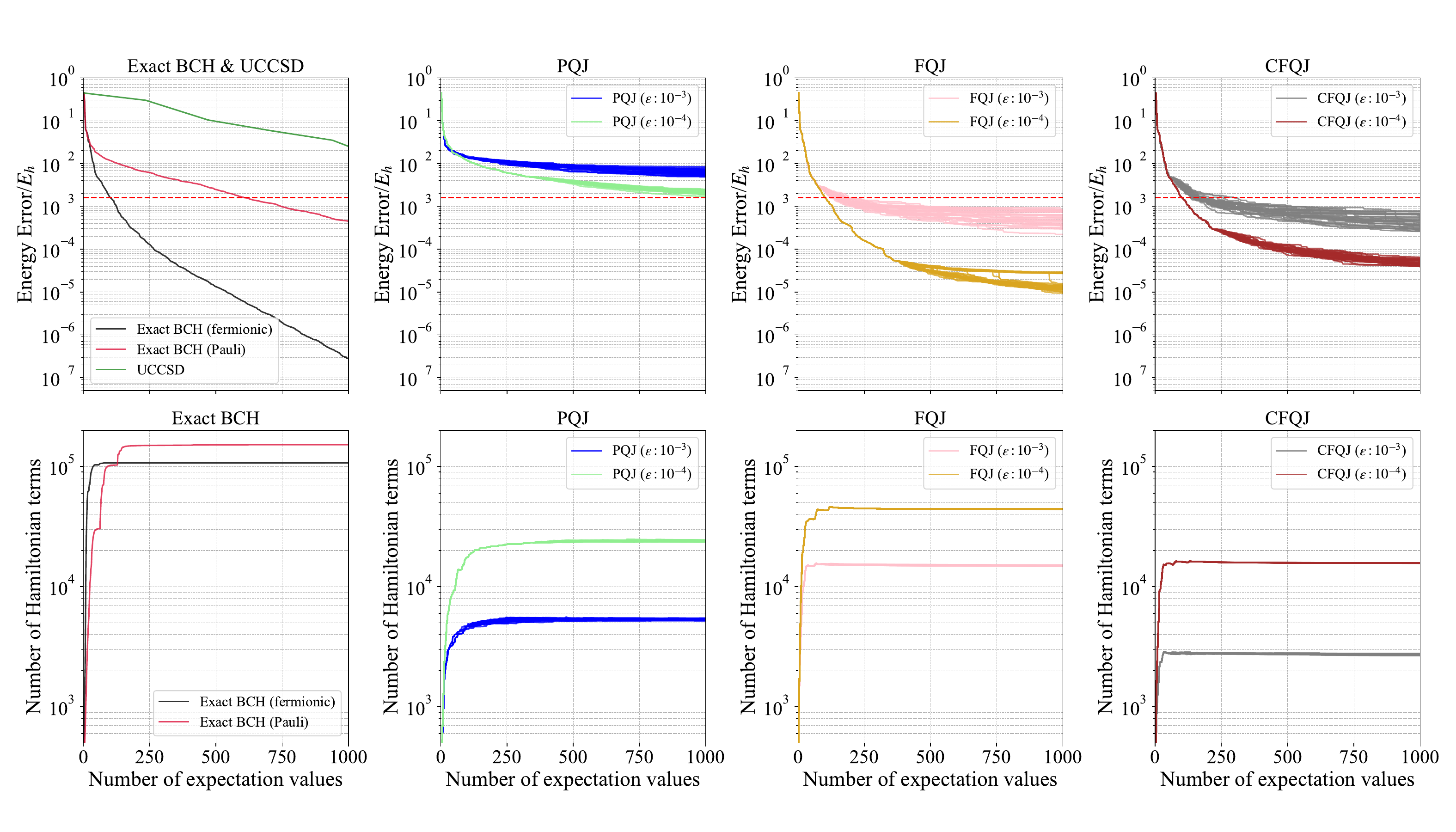}
  \caption{Energy error relative to FCI (top panels) and Hamiltonian term count (bottom panels) as a function of the number of expectation values for N$_2$ at the stretched bond length of 1.8~\AA. Results are shown for PQJ, FQJ, and CFQJ at truncation thresholds $\epsilon = 10^{-3}$ and $10^{-4}$, with comparisons to the exact BCH expansion and UCCSD. The red dashed lines in the top panels denote the chemical accuracy threshold.}
  \label{N2Eerror_1.8}
\end{figure*}

\subsection{Noiseless performance}
\label{sec:Noise-free}

\subsubsection{Energy convergence and Hamiltonian growth: N$_2$}
\label{sec:N2benchmark}

We focus on the nitrogen molecule (N$_2$) in a minimal STO-6G basis to investigate the behavior of the quantum Jacobi methods, examining energy convergence and Hamiltonian compression across different truncation schemes and operator bases. Calculations are performed within a (6e,6o) active space, obtained by restricting the system to the most relevant valence HF orbitals.
Two representative bond lengths are studied: 1.0977~\AA{} (Fig.~\ref{N2Eerror_1.0977}) and 1.8~\AA{} (Fig.~\ref{N2Eerror_1.8}), allowing us to probe both weakly and strongly correlated regimes. In both cases, the initial Hamiltonian consists of 246 fermion operators (or 246 Pauli operators after the Jordan-Wigner transformation). When truncation is applied, it is controlled by a coefficient threshold $\epsilon$ (varied between $10^{-3}$ and $10^{-4}$), with the cumulant approximation threshold set to $\kappa = 10\epsilon$. We perform 30 independent runs to capture variability arising from stochastic operator selection. VQE-UCCSD is included as a reference method and is implemented without enforcing point-group or spin symmetries. While the convergence behavior of VQE-UCCSD can depend on the choice of optimizer, we employ the L-BFGS algorithm~\cite{liu_limited_1989} for all calculations reported here. To compute first derivatives in VQE, we used the fermionic shift rule~\cite{kottmann_feasible_2021}, which requires estimating $2N_{\rm param}$ expectation values for a real wave function, in addition to the energy expectation value.

Figs.~\ref{N2Eerror_1.0977} and \ref{N2Eerror_1.8} compare the performance of Jacobi-based diagonalization methods at equilibrium (1.0977~\AA) and stretched (1.8~\AA) geometries of \(\mathrm{N}_2\). The upper panels report the energy deviation from the FCI reference, defined as \(\big|E_{\text{method}} - E_{\text{FCI}}\big|\) (in Hartree), while the lower panels show the number of Hamiltonian terms present in the Hamiltonian at each cycle. Both quantities are plotted as a function of the cumulative number of measured expectation values, with two expectation values collected per Jacobi cycle (Eqs.~(\ref{eq:Ekmu}–\ref{eq:cmu})).
The remaining four approaches are based on the QJ framework: an untruncated fermionic formulation using the exact BCH expansion, a Pauli-encoded truncated variant (PQJ), a truncated fermionic variant (FQJ), and a cumulant-decomposed extension of the latter (CFQJ).
\begin{table}[t]
\centering
\renewcommand{\arraystretch}{1.3}
\begin{tabular}{|c|c|c|c|c|}
\hline
\textbf{Geometry} & $\boldsymbol{\epsilon}$ & \textbf{PQJ} & \textbf{FQJ} & \textbf{CFQJ} \\
\hline
\multirow{2}{*}{1.0977\,\AA} 
    & \hspace{6pt}$10^{-3}$\hspace{6pt} & \hspace{6pt}90\hspace{6pt} & \hspace{6pt}90\hspace{6pt} & \hspace{6pt}114\hspace{6pt} \\
\cline{2-5}
    & \hspace{6pt}$10^{-4}$\hspace{6pt} & \hspace{6pt}262\hspace{6pt} & \hspace{6pt}160\hspace{6pt} & \hspace{6pt}154\hspace{6pt} \\
\hline
\multirow{2}{*}{1.8\,\AA} 
    & \hspace{6pt}$10^{-3}$\hspace{6pt} & \hspace{6pt}102\hspace{6pt} & \hspace{6pt}78\hspace{6pt} & \hspace{6pt}54\hspace{6pt} \\
\cline{2-5}
    & \hspace{6pt}$10^{-4}$\hspace{6pt} & \hspace{6pt}346\hspace{6pt} & \hspace{6pt}376\hspace{6pt} & \hspace{6pt}218\hspace{6pt} \\
\hline
\end{tabular}
\caption{Expectation values $2k_{\rm c}$ at which each Jacobi variant switches from deterministic to stochastic selection. Columns list the molecular geometry, truncation threshold $\epsilon$, and values for PQJ, FQJ, and CFQJ. These mark the onset of curve splitting in Figs.~\ref{N2Eerror_1.0977}--\ref{N2Eerror_1.8}.}

\label{tab:curve_split}
\end{table}
In both geometries, the exact fermionic and Pauli BCH expansions provide useful baselines. The fermionic variant converges faster because the rotated Hamiltonian preserves particle number, unlike the Pauli variant. Although both ultimately achieve chemical accuracy, this comes at the cost of uncontrolled operator growth: the transformed Hamiltonian expands to more than \(1.0\times 10^5 - 1.5\times 10^5\) terms. This highlights how rapidly the exact BCH expansion grows; we observe that, for both geometries, the term count saturates after the evaluation of roughly 100 expectation values, after which it remains essentially constant even though the energy continues to decrease.
The energy-error trajectories reveal marked contrasts between the near-equilibrium and strongly correlated regimes.

In contrast, QJ methods generally produce sparser Hamiltonians, particularly when a larger threshold $\epsilon$ and/or the cumulant approximation is used. The trade-off is that the resulting Hamiltonian is only a classical approximation, so the generator selection starts to deviate from the exact choice obtained by a quantum computer. When the deterministic selection scheme suggests the same generator as in the previous cycle (the critical cycle $k_{\rm c}$), we switch to the stochastic selection scheme. Consequently, the curves in Figs.~\ref{N2Eerror_1.0977}--\ref{N2Eerror_1.8} split beyond $k_{\rm c}$ (or $2k_{\rm c}$ expectation values), reflecting the 30 independent stochastic samples. Table 1 summarizes the values of $k_{\rm c}$ for different conditions.
As expected, introducing more approximations generally leads to an earlier transition from deterministic to stochastic selection.

At near-equilibrium geometry (top panels of Fig.~\ref{N2Eerror_1.0977}), PQJ converges smoothly but more slowly, whereas FQJ reaches chemical accuracy earlier, with intermittent sharp drops in error that further reduce the energy toward the FCI limit. These discontinuities reflect the stochastic nature of operator selection. CFQJ closely tracks FQJ in accuracy while maintaining a consistently smaller operator count, making it the most balanced approach overall. 
At stretched geometry (top panels of Fig.~\ref{N2Eerror_1.8}), PQJ does not reach chemical accuracy within \(10^3\) expectation values, reflecting the intrinsic limitations of Pauli-based truncation in multireference regimes. Nevertheless, PQJ consistently yields lower energy errors than UCCSD across both geometries, highlighting its superior convergence behavior. The fermionic variants achieve errors below 1.6~m$E_\mathrm{h}$ even in this challenging case, albeit at the cost of larger Hamiltonian expansions compared to equilibrium conditions.  

\begin{figure}
  \centering
  \includegraphics[width=0.5\textwidth]{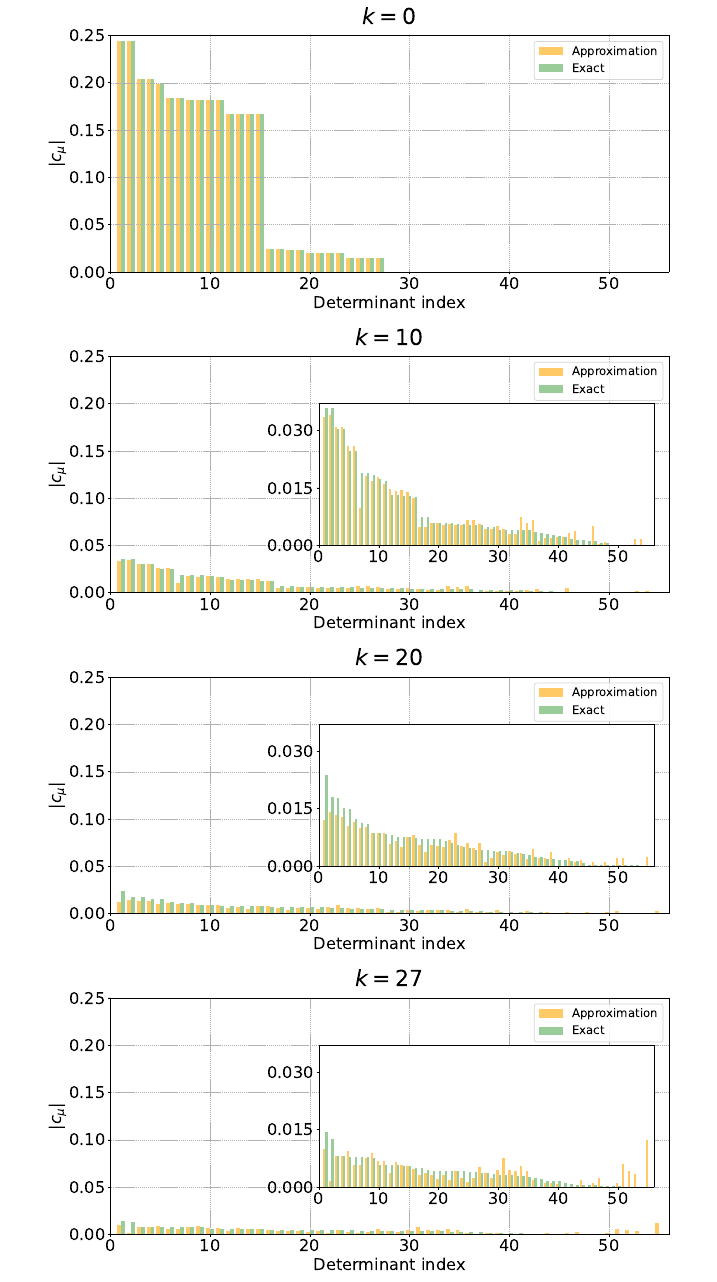}
  \caption{Residual excitation amplitudes for CFQJ method with a truncation threshold of $10^{-3}$ for N$_2$ at a bond length of 1.8~\AA. Each panel corresponds to a different iteration cycle ($k = 0,10,20$ and $k_c=27$), with insets providing zoomed views of selected regions for clarity.}
  \label{fig:k_amplitudes}
\end{figure}

The operator-count analysis (bottom panels of Figs.~\ref{N2Eerror_1.0977}--\ref{N2Eerror_1.8}) emphasizes the resource advantages of truncation-based schemes, which systematically suppress low-amplitude contributions. FQJ reaches sub-chemical precision for thresholds in the range $\epsilon=10^{-3}$--$10^{-4}$. CFQJ further enhances compression: at $\epsilon=10^{-4}$, it yields a Hamiltonian of comparable size to FQJ at $\epsilon=10^{-3}$, while maintaining similar performance. This efficiency gain stems from the cumulant decomposition, which eliminates negligible contributions without significantly compromising accuracy.

\subsubsection{Residual amplitude distribution: N$_2$}  
\label{sec:Residual amplitude distribution.}

Further insight into the convergence behavior can be gained by analyzing the residual amplitudes $c_\mu^{(k)}$ defined in Eq.~(\ref{eq:residual}) and its approximation $\tilde c_\mu^{(k)}$ defined in Eq.~(\ref{eq:approximate_r}). Fig.~\ref{fig:k_amplitudes} compares their distributions obtained with CFQJ during the deterministic operator-selection stage for Jacobi iterations \(k = 0, 10, 20\) and at the final deterministic iteration \(k_c=27\). Absolute amplitudes are shown for determinants sorted in descending order of exact absolute amplitude, including only those with \(|c_\mu| > 10^{-6}\). 

\begin{figure*}[t]  
  \includegraphics[width=\textwidth]{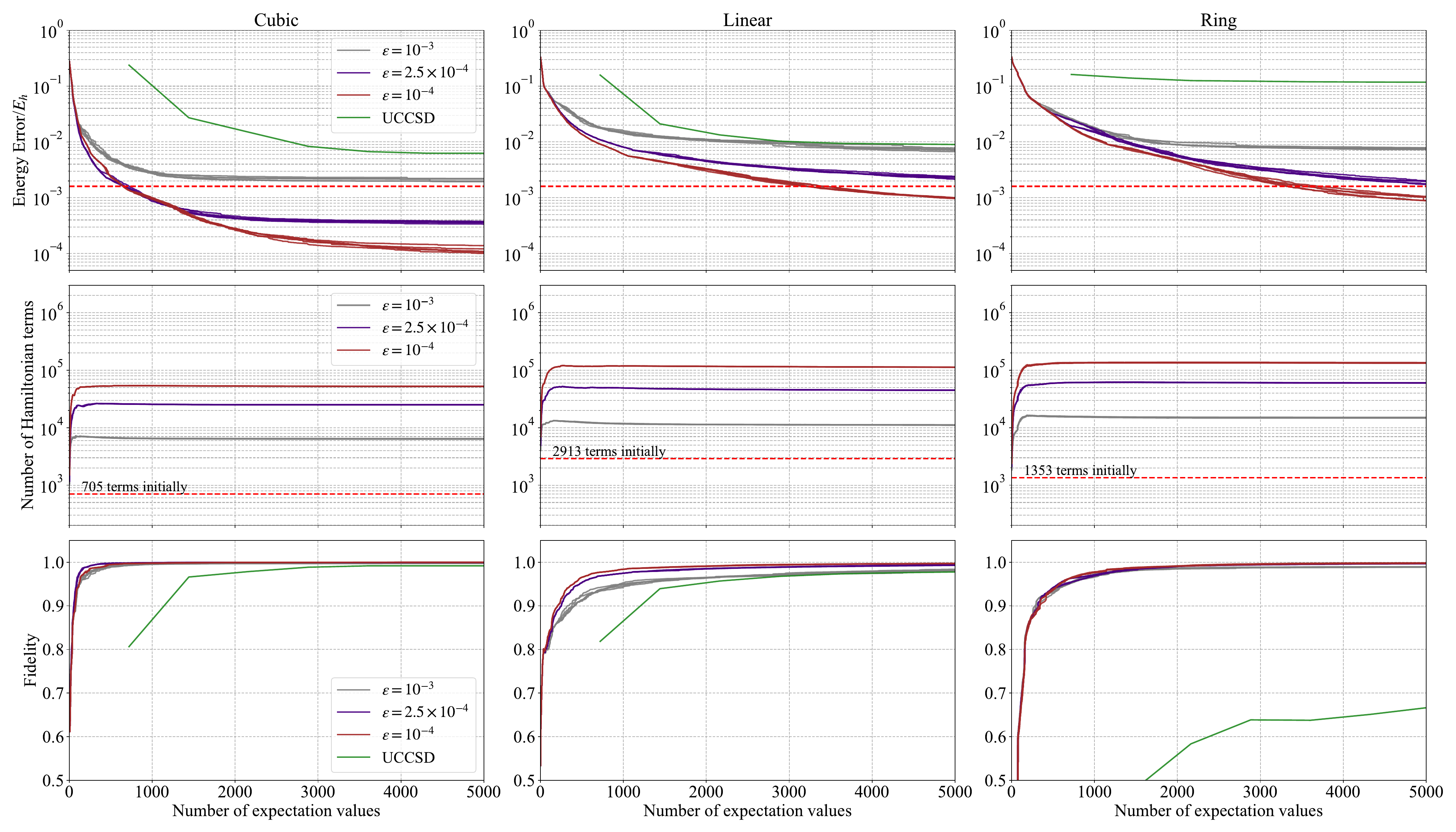}
  \caption{
Energy error relative to FCI (top row), number of Hamiltonian terms (second row) and fidelity with respect to the FCI state (third row) as functions of the number of expectation values for H$_8$ at a bond length of 1.5~\AA\ in cubic, linear, and ring geometries. Results are shown for CFQJ with truncation thresholds $\epsilon = 10^{-3},\, 2.5\times10^{-4},\, 10^{-4}$ and for UCCSD. Red dashed lines indicate chemical accuracy in the energy panels; dashed lines in the Hamiltonian panels denote the initial number of Hamiltonian terms.
}
  \label{H8error_1.5}
\end{figure*}

At the initial cycle ($k = 0$), the excitation distribution is narrow, comprising only singles and doubles,  with peak amplitudes near 0.25. As higher-order excitations are introduced, the distribution broadens.
This broadening, however, is accompanied by rapid amplitude suppression: within just a few cycles ($k = 10$), the largest amplitudes fall below 0.04. From this point onward, no determinant exceeds this threshold, highlighting the efficiency of the selection strategy in filtering dominant contributions early.
As a result, by early cycles ($k \le 10$), both the approximate and exact distributions are already broad, with several determinants contributing with comparable weights.
As iterations continue, the overall residual amplitudes become monotonically reduced. In particular, large amplitudes are strongly suppressed, whereas many small amplitudes persist, reflecting dynamical correlation.
From $k = 20$, discrepancies begin to emerge as the approximate residuals occasionally prioritize large-amplitude determinants that are not dominant in the exact distribution. At $k=27$, the largest $\tilde c_{\mu}$ is outside the excitation manifold of the exact BCH expansion due to the approximation employed. At this point, the deterministic algorithm becomes trapped with this outlier, because the actual $c_\mu$ evaluated with a quantum computer is zero or negligible, resulting in no net update on the Hamiltonian. This illustrates a limitation of a purely residual-dominant strategy in strongly correlated regimes, where broader exploration of excitations becomes necessary.
Nonetheless, if the population of such outlier determinants remains small and the overall distribution is otherwise similar to the exact one, the Monte Carlo sampling becomes effective. By broadening the exploration beyond residual-dominant excitations, the stochastic step increases the likelihood of selecting determinants that, although associated with smaller residual amplitudes, can still contribute appreciably to the energy. This probabilistic exploration is particularly relevant in multi-reference situations, where non-perturbative correlation is distributed over many comparable excitations.

\black
\subsubsection{Scalability and point group dependence: H$_8$}
\label{sec:H8}

We next consider the $\mathrm{H}_8$ system as a stringent test of the scalability and structural robustness of the CFQJ protocol.
As an eight-atom hydrogen cluster with pronounced multireference character, $\mathrm{H}_8$ admits linear, ring, and cubic geometries that differ in connectivity and dimensionality while sharing the same number of electrons and orbitals.
This setting enables us to isolate how dimensionality-induced correlation patterns influence algorithmic efficiency, Hamiltonian compression, and operator growth.

A clear geometry-dependent trend emerges in Fig.~\ref{H8error_1.5}, where three truncation thresholds, $\epsilon = 1.0\times 10^{-3}$, $2.5\times 10^{-4}$, and $1.0\times 10^{-4}$, are employed, each with ten independent runs.
Across all geometries, FQJ converges significantly faster than UCCSD in terms of both energy error and fidelity.
However, the convergence rate strongly depends on the underlying structure.
The cubic geometry exhibits the most rapid convergence, reaching chemical accuracy with substantially fewer expectation values than the linear and ring configurations.
This accelerated reduction in energy error is accompanied by a correspondingly rapid increase in fidelity, indicating a consistent improvement in both energetic and wave function accuracy.
In contrast, the linear and ring geometries converge more slowly under identical truncation thresholds, with the ring configuration showing the most delayed approach to chemical accuracy at tighter $\epsilon$ values.
Importantly, while UCCSD fails to achieve high fidelity for the strongly correlated ring geometry, FQJ consistently attains near-unity fidelity across all structures, demonstrating its robustness in challenging correlation regimes.

To elucidate the origin of this geometry-dependent convergence behavior, we analyze the residual amplitude distributions obtained with $\epsilon = 10^{-4}$.
While the preceding discussion focused on energy convergence and operator growth, the residual vector provides direct insight into how correlation weight is distributed across the excitation manifold during the iterative procedure.
By examining the statistical structure of this distribution, one can distinguish between regimes dominated by a small number of large components and those characterized by broad delocalization over many operators.

\begin{figure}[t]  
  \centering
  \includegraphics[width=0.45\textwidth]{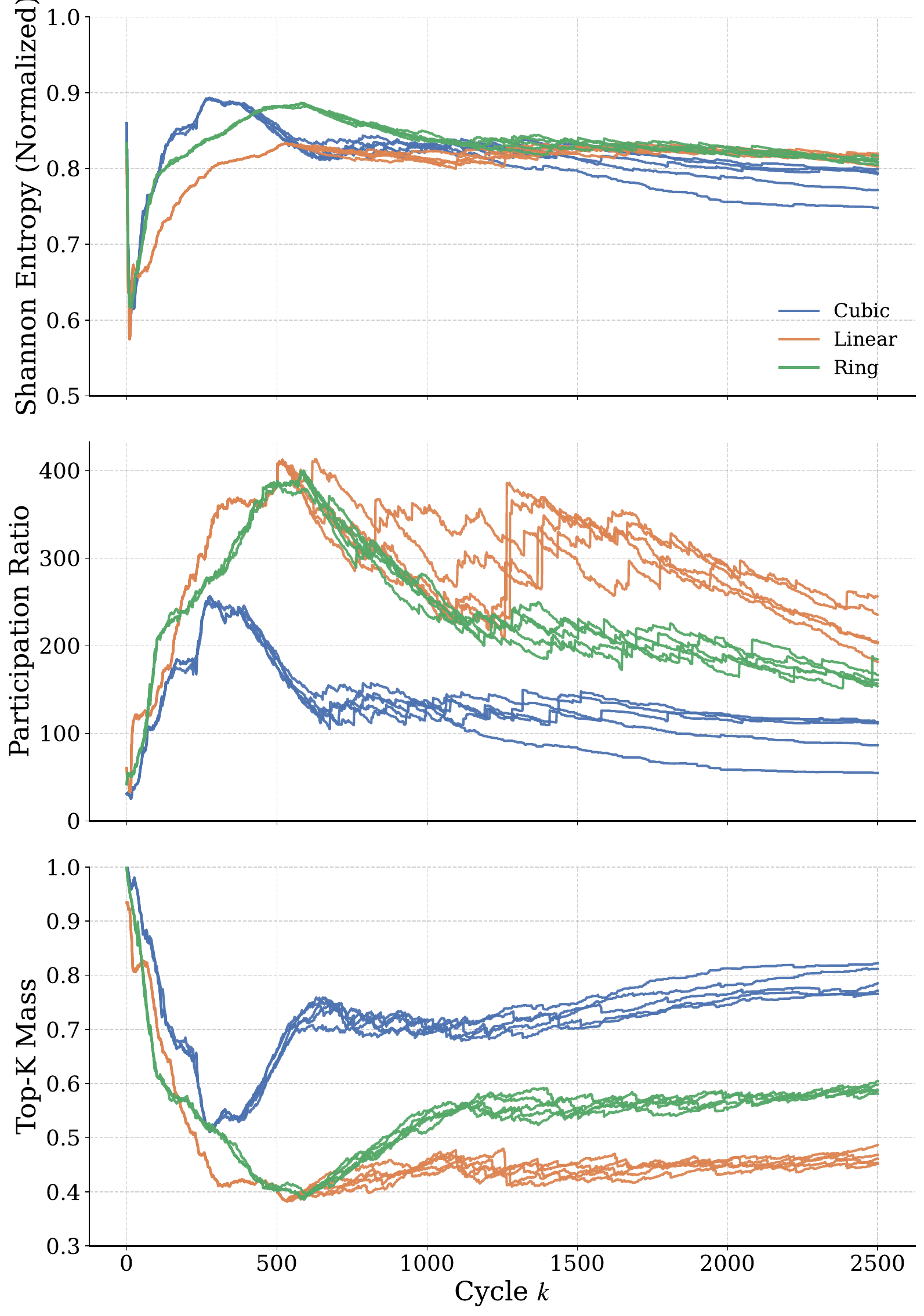}
  \caption{
Evolution of residual elements from CFQJ at fixed threshold $\epsilon=10^{-4}$.
Shown are the normalized Shannon entropy (top), the participation ratio (middle), and the top-$K$ probability mass with $K=100$ (bottom), as functions of the iteration cycle $k$.
Results are presented for the cubic, linear, and ring geometries of the H$_8$ system.
}
  \label{H8_entropy}
\end{figure}

To quantify this behavior, we introduce the normalized Shannon entropy,
\begin{align}
S^{(k)} =
\frac{
-\displaystyle\sum_{\mu=1}^{N} p_\mu^{(k)} \ln p_\mu^{(k)}
}{
\ln N
},
\label{eq:shannon_entropy}
\end{align}
where the normalized weights are defined as $
p_\mu^{(k)} = |c_\mu^{(k)}|^2$ (analogous to  Eq.~(\ref{eq:proba_dist}))
and $N$ denotes the number of residual components exceeding $10^{-12}$ in magnitude.
As shown in the top panel of Fig.~\ref{H8_entropy}, $S^{(k)}$ increases during the early iterations and eventually saturates for all geometries.
This saturation indicates that, as the algorithm progresses, the residual weight becomes distributed among many operators of comparable magnitude, a signature of predominantly dynamical correlation.

Because entropy alone does not fully resolve the geometric differences, we further consider the participation ratio (PR),
\begin{equation}
\mathrm{PR}^{(k)} =
\frac{1}{\displaystyle\sum_{\mu=1}^{N} \left(p_\mu^{(k)}\right)^2},
\label{eq:participation_ratio}
\end{equation}
which estimates the effective number of significant contributors to the residual distribution.
We also define the Top-$K$ mass as
\begin{equation}
M_K^{(k)} =
\sum_{i=1}^{K} p_{(i)}^{(k)},
\label{eq:topk_mass}
\end{equation}
where $p_{(i)}^{(k)}$ are the weights sorted in descending order.
This quantity measures the fraction of the total residual weight captured by the $K$ largest components.

\begin{figure*}[t]
\centering
\includegraphics[width=0.9\textwidth]{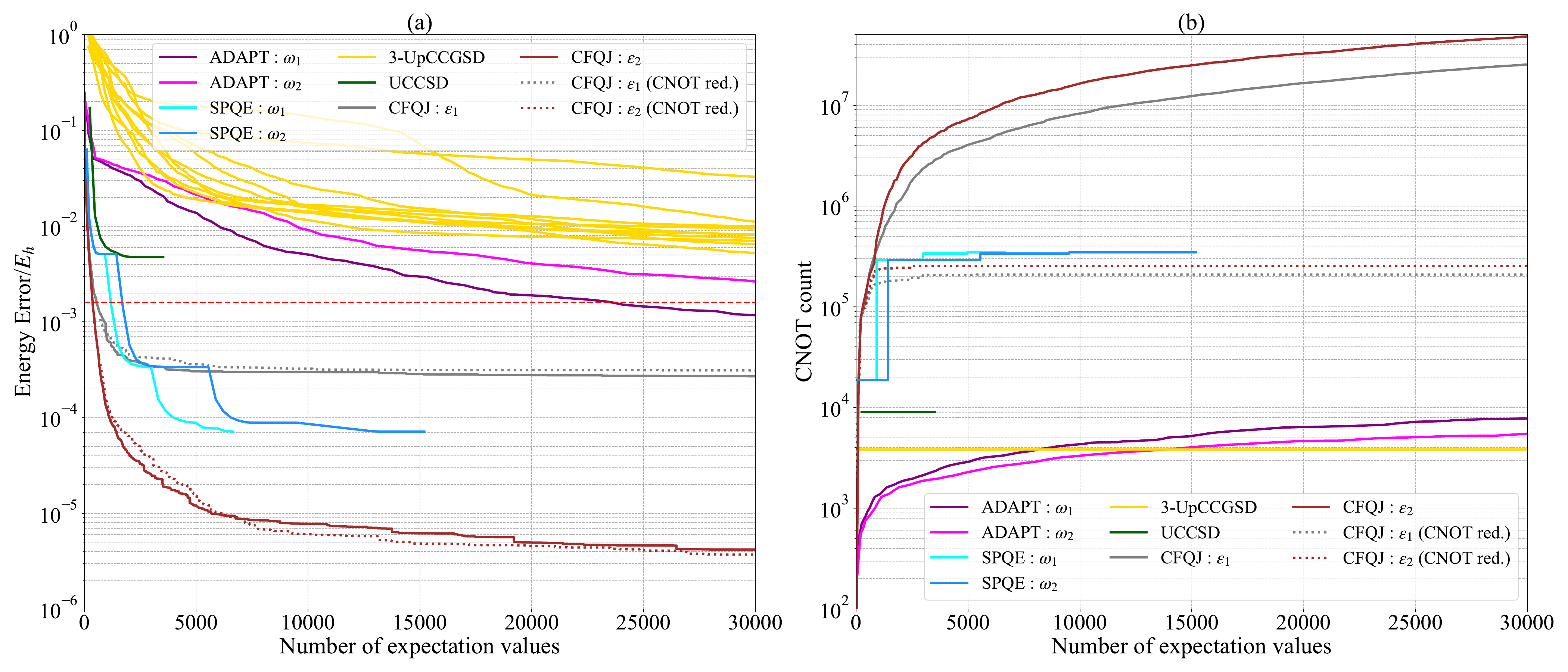}
\caption{
(a) Energy error relative to FCI as a function of the number of expectation values
for H6 at a bond length of 1.5~\AA. (b) Circuit complexity measured in CNOT count.
We compare CFQJ with truncation thresholds $\epsilon_1 = 5\times10^{-4}$ and $\epsilon_2 = 5\times10^{-5}$, including a dynamical gate reduction variant in which rotations with angles $|\theta|<10^{-2}$ are absorbed into previously applied operators. Results are also shown for ADAPT-VQE in the spin-adapted formulation ($\omega_1 = 10^{-3}$ and $\omega_2 = 10^{-5}$), UCCSD, $k$-UpCCGSD with $k=3$ layers, and SPQE with macro-iteration threshold $\Omega = 10^{-2}$, time step $\Delta t = 0.1$, and micro-iteration thresholds ($\omega_1, \omega_2$).
}
\label{H6_benchmark}
\end{figure*}

The evolution of $\mathrm{PR}^{(k)}$ and $M_{K=100}^{(k)}$ reveals a clear hierarchy among the geometries.
For the cubic structure, $\mathrm{PR}^{(k)}$ initially increases but subsequently decreases and stabilizes around $\mathrm{PR} \approx 100$, while $M_{100}^{(k)}$ exceeds $70\%$, indicating that a relatively small subset of operators carries the majority of the residual weight during the later stages of the iteration.
In contrast, the ring geometry maintains a substantially larger participation ratio throughout the iteration, with $M_{100}^{(k)}$ remaining in the range of $50$--$60\%$.
The linear geometry exhibits the most diffuse distribution, characterized by the largest $\mathrm{PR}^{(k)}$ and a Top-100 mass below $50\%$.

Thus, although the entropy values appear similar across geometries, the participation ratio and Top-$K$ mass clearly demonstrate that the cubic system possesses the most localized residual structure, whereas the linear system exhibits the strongest delocalization across operator space.

This residual localization directly explains the truncation sensitivity observed in Fig.~\ref{H8error_1.5}.
In the cubic geometry, where the residual weight is concentrated in a relatively small subset of operators, aggressive truncation primarily removes negligible components and therefore preserves rapid convergence.
In contrast, for the ring and linear geometries, where the residual weight is broadly distributed over many moderately sized operators, larger truncation thresholds ($\epsilon = 2.5\times 10^{-4}$ and $1.0\times 10^{-3}$) eliminate a non-negligible fraction of physically relevant contributions, resulting in slower convergence.

Overall, these results indicate that the convergence behavior of CFQJ is governed not solely by the overall correlation strength, but by the distributional structure of the residual amplitudes.
Geometries exhibiting localized residual weight favor efficient Hamiltonian compression and rapid convergence, whereas systems with broadly delocalized residual contributions present a more challenging scenario for adaptive operator selection.  Nevertheless, given that FQJ attains high fidelity across all geometries, the remaining correlation effects are expected to be predominantly dynamical in nature. Such residual dynamical correlation, being broadly distributed yet relatively small in magnitude, should be amenable to subsequent perturbative treatment, suggesting a natural route toward hybrid or post-FQJ correction schemes.

\subsection{Circuit complexity in a strongly correlated H$_6$ chain}  
\label{sec:CircuitcomplexityH6}

While QJ is primarily designed for initial state preparation in FTQC, reducing its circuit depth is highly desirable in practice. A naive implementation requires repeated additions of quantum gates at each cycle, just like QITE,\cite{motta_determining_2020,tsuchimochi_multi-state_2023,tsuchimochi_improved_2023,gomes_efficient_2020} leading to a continual increase in circuit depth. Here, we therefore examine the gate-reduction scheme introduced in Section~\ref{sec:CNOT} to assess its effectiveness in reducing CNOT gate counts and the resulting trade-off in energy accuracy.

To this end, we consider a linear hydrogen chain (H$_6$) as a benchmark system. 
This system provides a suitable balance between physical complexity and computational tractability, allowing a systematic evaluation of both accuracy and circuit cost across methods. Using this model, we compare  energy convergence behavior and the growth of CNOT counts  among the representative VQE methods including UCCSD, ADAPT-VQE\cite{grimsley_adaptive_2019} and $k$-UpCCGSD,\cite{lee_generalized_2019} and also SPQE.\cite{EvangelistaPQE_2021} To enable a fair comparison across methods, convergence is quantified in terms of the cumulative number of measured expectation values rather than the number of variational parameters.

For ADAPT-VQE, we used a spin-unrestricted operator pool $\{\hat A_i\}$ consisting of generalized singles and doubles\cite{Tsuchimochi_adapt_2022}. At each macro-iteration, the operator with the largest gradient magnitude within the pool is selected and its parametrized exponential is added to the ans\"atze. In principle, this procedure requires evaluation of the three-body reduced density matrix for computing the commutator $[\hat H, \hat A_i]$, which introduces additional measurement overhead. However, to enable a direct comparison between methods based solely on the number of expectation value measurements, this extra cost was not included in our analysis. In practice, we consider this overhead negligible compared with the measurement cost associated with the gradient evaluations required for VQE energy minimization
For each VQE optimization, we employed the BFGS algorithm and show the results for two gradient-norm convergence thresholds, $\omega_1 = 10^{-3}$ and $\omega_2 = 10^{-5}$. 

In $k$-UpCCGSD, we used $k=3$ layers and initialized the parameters randomly, as its energy landscape contains many local minima; ten independent optimization runs are shown to illustrate the variability in convergence behavior. 
SPQE calculations employed a macro-iteration residual-norm threshold $\Omega = 10^{-2}$ and a time step of $\Delta t = 0.1$, while the micro-iteration threshold was set to the same values as the ADAPT-VQE gradient-norm convergence thresholds, namely, $\omega_1 = 10^{-3}$ and $\omega_2 = 10^{-5}$. As in ADAPT-VQE,  the measurement cost for operator selection based on the residual equation (Eq.~\ref{Residual_SPQE}) was neglected. The CNOT count for SPQE was estimated as twice that required for implementing $\hat U$ and the short time evolution (without Trotterization), which is needed for operator selection through Eq.~(\ref{Residual_SPQE}).

Within the CFQJ framework, we employ two Hamiltonian-term truncation thresholds, $\epsilon_1 = 5\times10^{-4}$ and $\epsilon_2 = 5\times10^{-5}$, allowing us to quantify the trade-off between accuracy and circuit-depth induced by Hamiltonian compression.

Figure~\ref{H6_benchmark} summarizes the trade-off between measurement overhead and circuit depth. As can be seen in the left panel, UCCSD and 3-UpCCGSD fail to reach chemical accuracy within the explored measurement range, highlighting the limitations of fixed single-reference ans\"atze.
\begin{figure*}[t] 
  \centering
  \includegraphics[width=0.8\textwidth]{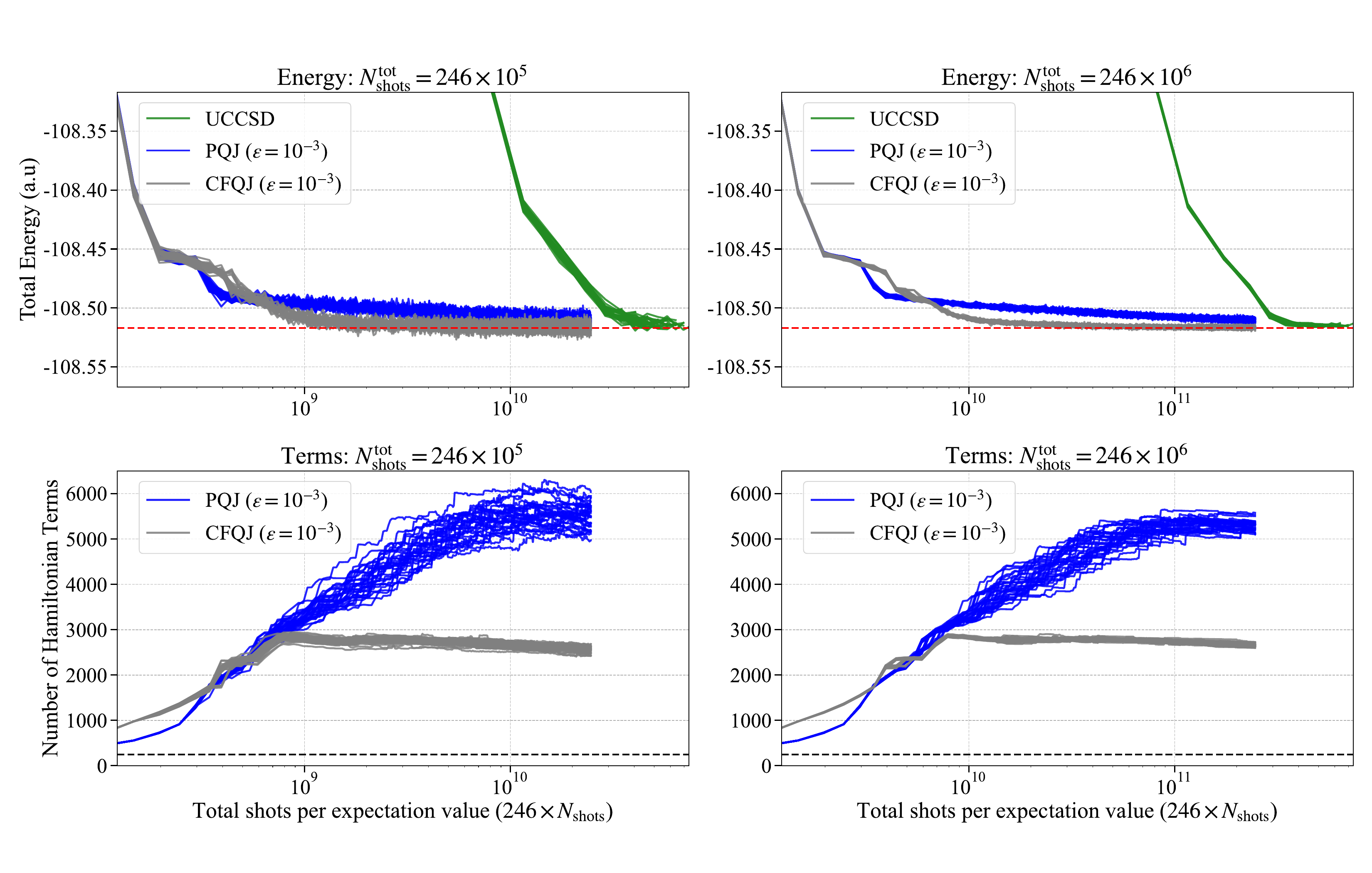}
  \caption{Evolution of the total energy and number of Hamiltonian terms for N$_2$ at 1.8~\AA{} under shot noise with $N_{\text{shots}} = 10^5$ and $10^6$ per term ($N_{\text{shots}}^{\mathrm{tot}} = 246 \times N_{\text{shots}}$ per expectation value). The red and black dashed lines indicate the chemical accuracy and the initial number of Hamiltonian terms, respectively.}
  \label{N2shots_1.8_eig}
\end{figure*}
In contrast, ADAPT-VQE, SPQE, and CFQJ all achieve chemical accuracy. However, ADAPT-VQE requires more than $2\times10^{4}$ expectation values (depending on the gradient threshold), whereas both SPQE and CFQJ converge within fewer than $2.5\times10^{3}$ measurements.
This favorable measurement scaling of CFQJ comes at a substantial circuit cost. In its original uncompressed gate form, CFQJ produces circuits exceeding $10^{7}$ CNOT gates (right panel), for both Hamiltonian truncation thresholds $\epsilon_1$ and $\epsilon_2$. Although convergence is already obtained with the looser threshold $\epsilon_1$, the circuit depth is dominated not by the final Hamiltonian size but by the cumulative sequence of Pauli rotations selected throughout the iterative procedure. As a result, the initial implementation becomes impractical from a circuit perspective. In contrast, merging the gates reduces the CNOT count by several orders of magnitude while preserving the energy curves within the statistical fluctuations. The circuit depth is thereby brought down to the same order of magnitude as SPQE, enabling a meaningful circuit-level comparison. This accuracy–complexity interplay, controlled by the angle threshold used in the gate-merging step, is expected to yield similar circuits to SPQE~\cite{EvangelistaPQE_2021}. However, despite this substantial reduction, the resulting circuit depth remains markedly higher than that of UCCSD, ADAPT-VQE, and 3-UpCCGSD, whose ans\"atze yield substantially shallower circuits, as was previously observed in SPQE\cite{EvangelistaPQE_2021}.

Together, these results expose a clear trade-off between measurement overhead and circuit depth. CFQJ shifts the balance towards measurement efficiency and, when combined with dynamical circuit compression, attains a circuit depth that may be viable within fault-tolerant settings. The remaining circuit depth originates from the broader excitation manifold explored by CFQJ, including higher-body operators beyond conventional doubles. 
Although this expanded operator manifold improves convergence in energy, it inevitably translates into a greater number of correlated excitation operators to be realized on the quantum circuit. This highlights a fundamental compromise between compact excitation manifolds and measurement efficiency.

\black

\subsection{Effect of Finite Sampling Noise: N$_2$}
\label{sec:N-shots}

In practical quantum simulations, shot noise arising from finite sampling of expectation values can strongly influence both accuracy and resource requirements. It is therefore essential to evaluate how different algorithmic variants behave under realistic shot noise. To this end, we benchmark VQE-UCCSD, PQJ, and CFQJ at two representative shot budgets, $N_{\text{shots}} = 10^5$ and $10^6$ where $N_{\text{shots}}$ denotes the number of samples per term for all 246 terms in $\hat H$. Here, we will not attempt to reduce measurement costs with strategies such as commuting groups\cite{commutingroup_2024} or unitary partitioning\cite{izmaylov_unitary_2020,Measurement_reduction2022}.
Fig.~\ref{N2shots_1.8_eig} summarizes these results with the total energy (top panels) and the number of Hamiltonian terms (bottom panels) as a function of the number of shots, for the N$_2$ molecule at a bond length of 1.8~\AA{}.

Across both noise levels, CFQJ and PQJ consistently outperform UCCSD in terms of convergence speed. Both QJ methods eventually approach chemical accuracy with sufficient measurements, although their pathway vary significantly. CFQJ shows the most favorable behavior: even in the low-shot regime ($N_{\text{shots}} = 10^5$), its energy estimates remain close to the noiseless reference, with fluctuations that are both smaller in magnitude and less persistent across independent runs. PQJ follows a similar trend, though with slightly larger variance. Conversely, UCCSD requires significantly more measurements to approach chemical accuracy. This slower convergence may be due to the full variational optimization in the method, where the statistical noise in the gradient evaluations hinders the optimizer from reliably approaching the energy minimum.
The superior performance of Jacobi-based approaches can be traced back to their structural properties. Namely, they avoid expending effort on negligible directions by updating the most important excitation at each step to variationally lower the energy.
The bottom panels of Fig.~\ref{N2shots_1.8_eig} further reveals the evolution of the effective Hamiltonian size as measurements accumulate. While energy from both QJ methods remains relatively stable, the number of terms in the PQJ Hamiltonian exhibits more pronounced fluctuations compared to CFQJ. 
Nevertheless, both PQJ and CFQJ maintain a substantially reduced number of terms, with CFQJ showing the most aggressive compression.
Importantly, this compression is achieved without compromising accuracy, underscoring its dual advantage of lowering measurement costs and mitigating hardware resource demands. Such a balance between statistical robustness, accuracy, and efficiency suggests that the fermionic Quantum Jacobi algorithms could offer practical advantages, where shot noise and limited sampling budgets pose significant constraints.

\section{Conclusion}
\label{sec:conclusion}

In this paper, we have introduced a hybrid quantum--classical diagonalization framework based on sequential Givens rotations, with the goal of reducing quantum measurement overhead by shifting key parts of the update procedure to classical preprocessing. Across the bond-dissociation regimes considered, the method exhibits rapid convergence near equilibrium and remains competitive in the more strongly correlated cases studied. Comparisons across operator representations highlight the advantages of fermionic and cumulant-decomposed formulations: while the Pauli-based representation can deteriorate in multi-reference regimes, fermionic and cumulant-based variants more systematically recover accurate ground-state energies within the benchmark set. In particular, CFQJ provides an effective compression mechanism, achieving comparable accuracy under stringent threshold settings while reducing the retained operator count by up to an order of magnitude in our benchmarks. 

Our results further quantify a practical trade-off between energy accuracy and Hamiltonian growth. Tightening truncation thresholds reduces the energy error but increases the number of retained terms, whereas looser filtering keeps the operator set compact at the expense of limiting progress toward chemical accuracy. This compromise defines a practical operating regime and motivates adaptive strategies that improve accuracy without a prohibitive increase in computational resources. 
Residual amplitude analyses further clarify the convergence behavior of the algorithm. Diagnostics based on participation ratio, entropy, and Top-$K$ mass statistics suggest that the residual space is often dominated by a relatively small subset of operators, with their relative importance depending on molecular geometry. These observations rationalize the efficiency of the operator-selection procedure and provide insight into how correlation effects are distributed across the explored operator manifold.

At the same time, the residual analysis exposes an operational limitation of purely deterministic operator selection under approximate Hamiltonian updates. Deterministic updates can drive efficient early convergence but may stagnate once outlier residual component dominate the ranking of candidate operators. In contrast, Monte Carlo sampling promotes broader exploration of excitations and facilitate continued error reduction beyond this regime.

At the circuit level, the principal cost of fermionic QJ variants is the depth associated with accumulating many rotations. We therefore introduced a gate-reduction procedure that merges repeated generators with small angles, reducing circuit depth by orders of magnitude while maintaining energies within statistical uncertainty. 
Even after compression, the circuit depth remains comparatively deeper than typical NISQ-oriented ansatz methods underscoring that the most natural application domain is early fault-tolerant hardware where deeper circuits are feasible but measurement remains a limiting factor. Taken together, these findings highlight a three-way balance among measurement cost, energy precision, and circuit depth, and identify circuit compression as a necessary ingredient for scalable diagonalization-based workflows. 

Finally, simulations with finite sampling indicate that the framework remains robust under shot noise, with CFQJ showing enhanced stability to statistical fluctuations while still preserving substantial operator compression.

Overall, the Quantum Jacobi framework emerges as a viable alternative to wave-function–optimization approaches for correlated molecular simulations, particularly as a provider of high-quality zeroth-order states with reduced measurement demands.
Future work includes systematic integration with post-processing (e.g., perturbative corrections) and fault-tolerant refinements (e.g., phase estimation), as well as extensions to excited states and more general, including non-Hermitian, Hamiltonians.

\section*{Acknowledgements}
This work was supported by the JST FOREST Program, Grant No. JPMJFR223U. B. M acknowledges financial support from the SIT scholarship at Shibaura Institute of Technology.

\section*{DATA AVAILABILITY} 
The data that supports the findings of this study are available from the corresponding author upon reasonable request. 

\section*{Competing interests}
The authors declare no competing financial interest.

\bibliography{jacobi.bib}
\end{document}